\documentclass[a4paper,11pt]{article}
\usepackage{jcappub} % for details on the use of the package, please see the JINST-author-manual
\usepackage{lineno}
\usepackage{dcolumn}% Align table columns on decimal point
\usepackage{verbatim}

\usepackage{breakurl}

\usepackage{lmodern}

\pdfoutput=1
\usepackage{graphicx}
\usepackage{bm}
\usepackage{amsmath}
\usepackage{amsfonts}
\usepackage{amssymb}
\usepackage{multirow}
\usepackage[capitalise]{cleveref}
\usepackage{acro}
\usepackage[utf8]{inputenc}

\usepackage{tikz}
\usepackage{pgfplots}

\crefname{figure}{Fig.}{Figs.}
\def\({\left(}
\def\){\right)}
\def\[{\left[}
\def\]{\right]}

\newcommand{\be}{{\begin{eqnarray}}}
\newcommand{\ee}{{\end{eqnarray}}}

\newcommand{\Beq}{\begin{align}}
\newcommand{\Eeq}{\end{align}}

\DeclareAcronym{2nd-order}{
	short = 2nd-order ,
	long  = second-order ,
  short-plural = s ,
}

\DeclareAcronym{1st-order}{
	short = 1st-order ,
	long  = first-order ,
  short-plural = s ,
}

\DeclareAcronym{LHAASO}{
	short = LHAASO ,
	long  = Large High Altitude Air Shower Observatory ,
  short-plural = s ,
}

\DeclareAcronym{LIV}{
	short = LIV ,
	long  = Lorentz invariance violation ,
  short-plural = s ,
}

\DeclareAcronym{KM2A}{
	short = KM2A ,
	long  = Kilometer Squared Array ,
  short-plural = s ,
}

\DeclareAcronym{WCDA}{
	short = WCDA ,
	long  = Water Cherenkov Detector Array ,
  short-plural = s ,
}

\begin{document}

\title{\boldmath Probing Lorentz Invariance Violation at High Energies Using LHAASO Observations of GRB221009A via DisCan Algorithm}

\author[a,c]{Yu-Chen Hua,}
\author[b,c]{Xiao-Jun Bi,}
\author[b,c]{Yu-Ming Yang,}
\author[b]{Peng-Fei Yin}

\affiliation[a]{Theoretical Physics Division, Institute of High Energy Physics, Chinese Academy of Sciences, Beijing 100049, China}

\affiliation[b]{State Key Laboratory of Particle Astrophysics, Institute of High Energy Physics, Chinese Academy of Sciences, Beijing 100049, China}

\affiliation[c]{School of Physical Sciences, University of Chinese Academy of Sciences, 
Beijing 100049, China}

\emailAdd{ychua@ihep.ac.cn}
\emailAdd{bixj@ihep.ac.cn}
\emailAdd{yangyuming@ihep.ac.cn}
\emailAdd{yinpf@ihep.ac.cn}

% Collaborations

%% [A] If main author
%% \collaboration{\includegraphics[height=17mm]{collabroation-logo}\\[6pt]
%%  XXX collaboration}

%% or
%% [B] If "on behalf of"
%% \collaboration[c]{on behalf of XXX collaboration}

% Authors
% The "\note" macro will give a warning: "Ignoring empty anchor...", you can safely ignore it.

%% [A] simple case: 2 authors, same institution
%% \author[1]{A. Uthor\note{Corresponding author.}}
%% \author{and A. Nother Author}
%% \affiliation{Institution,\\Address, Country}

%% or, e.g.
%% [B] more complex case: 4 authors, 3 institutions, 2 footnotes
%% \author[a,b]{F. Irst,\}
%% \author[c]{S. Econd,}
%% \author[a,1]{T. Hird\note{Also at Some University.}}
%% \author[c,1]{and Fourth}
%% \affiliation[a]{Institution_1,\\Address, Country}
%% \affiliation[b]{Institution_2,\\Address, Country}
%% \affiliation[c]{Institution_3,\\Address, Country}

% \author{A. Uthor}
% \affiliation{One University,\\
% some-street, Country}
% \affiliation{Another University,\\
% different-address, Country}

% % E-mail addresses: only for the corresponding author
% \emailAdd{first@one.univ}

\abstract{
The Lorentz invariance violation (LIV) predicted by some quantum gravity theories would manifest as an energy-dependent speed of light,  which may potentially distort the observed temporal profile of photons from astrophysical sources at cosmological distances. 
The dispersion cancellation (DisCan) algorithm offers a powerful methodology for investigating such effects by employing quantities such as Shannon entropy, which reflects the initial temporal characteristics. In this study, we apply the DisCan algorithm to search for LIV effects in the LHAASO observations of GRB 221009A, combining data from both the \ac{WCDA} and \ac{KM2A} detectors that collectively span an energy range of $\sim 0.2-13$ TeV. Our analysis accounts for the uncertainties from both energy resolution and temporal binning. We derive $95\%$ confidence level lower limits on the LIV energy scale of $E_{\rm{QG}}/10^{19}~\text{GeV}>21.1$ (13.8) for the first-order subluminal (superluminal) scenario, and $E_{\rm{QG}}/10^{11}~\text{GeV}> 14.9$ (13.7) for the second-order subluminal (superluminal) scenario.}

\maketitle

%\flushbottom

\section{Introduction}

The fundamental incompatibility between general relativity and quantum mechanics has motivated the development of various quantum gravity theories, many of which predict a potential violation of Lorentz invariance at energies near or above the Planck scale \cite{Kostelecky:1988zi, Amelino-Camelia:1996bln, Carroll_2001, Amelino-Camelia:2002cqb, Amelino_Camelia_2013, Addazi_2022}. Such \ac{LIV} could induce energy-dependent modifications to the vacuum dispersion relationship of particles, potentially altering the speed of light and producing observable experimental signatures. This renders the search for \ac{LIV} effects one of the most promising approaches to investigating quantum gravity phenomenology.

The manifestation of \ac{LIV} effects is particularly favorable at high energies and over cosmological distances, making high-energy photons from distant astrophysical sources ideal probes for \ac{LIV} signatures \cite{1999PhRvL..83.2108B, Albert_2008, Abramowski_2011,chrétien2015constraining,2017, Amelino_Camelia_1998, Albert_2020, Acciari_2020, LHAASO:2021opi, Terzic:2021rlx, Piran:2023xfg, Yang:2023kjq, LHAASO:2024lub, xi2025constraintslorentzinvarianceviolation}. Previous studies have employed various astrophysical detectors and analytical techniques to investigate different \ac{LIV} manifestations, including energy-dependent photon velocity \cite{1999PhRvL..83.2108B, Albert_2008, Abramowski_2011, chrétien2015constraining, 2017, Aharonian_2008, Ellis_2003, Vasileiou_2013, Mart_nez_2009}, photon decay or splitting \cite{LHAASO:2021opi, Albert_2020}, and modifications to pair production thresholds \cite{Abdalla_2019, Li_2023}, yielding stringent constraints on the \ac{LIV} energy scale.

The recent observation of GRB 221009A at redshift z = 0.151, the most luminous long gamma-ray burst, has provided good opportunities for new physics studies. In 2022, the \ac{LHAASO} detected numerous high-energy photon events ranging from 0.3 TeV to approximately 10 TeV using both its \ac{WCDA} and \ac{KM2A} \cite{2023_1, 2023_2} detectors, offering an excellent dataset to investigate potential LIV effects at these energies. Several investigations have already utilized these observations to set stringent constraints on LIV parameters \cite{Piran:2023xfg, Yang:2023kjq, LHAASO:2024lub, xi2025constraintslorentzinvarianceviolation}.

In this study, we employ the dispersion cancellation (DisCan) algorithm \cite{Scargle2008} to investigate possible LIV effects \cite{ Zitzer:2013gka, xi2025constraintslorentzinvarianceviolation} in the LHAASO observational data of GRB 221009A. If \ac{LIV} exits, high energy photons emitted from the source exhibit energy-dependent velocities during their propagation, potentially distorting the observed temporal profile of the burst.
We can reconstruct the time profile of photons reflecting the initial temporal properties of the emission at the source.
The correct LIV parameters should correspond to a reasonable initial time profile characterized by specific structures, which do not necessarily require prior knowledge for determination and can be evaluated using Shannon entropy \cite{Scargle2008}. We utilize the minimum Shannon entropy to explore potential LIV effects and establish constraints on the LIV parameters. 
%In \cite{xi2025constraintslorentzinvarianceviolation} a similar method is adopted to study LIV, however, the treatment of the Shannon entropy calculation in our study is quite different. 
In our analysis, we incorporate observations from both the \ac{WCDA} and \ac{KM2A} detectors, which span a wide energy range, where LIV effects may be substantially manifested. To set constraints on the LIV parameters, we
also take into account the impacts of energy resolution and temporal binning effects.

This paper is organized as follows. In \cref{sec:LIV-DisCan}, we provide a brief overview of the LIV effects and introduce the DisCan algorithm. We demonstrate its feasibility through simulated photon data.
In \cref{sec:mine}, we apply the DisCan algorithm to the \ac{LHAASO} observations of GRB221009A to constrain the \ac{LIV} parameters. We also consider the impacts of energy resolution and temporal binning effects on the results. 
In \cref{sec:Conclusion}, we provide a summary. 

\section{Lorentz invariance violation effects and the dispersion cancellation algorithm}\label{sec:LIV-DisCan} 

The conventional dispersion relation of photons in vacuum, expressed as $E^2 = p^2 c^2$, holds with high precision at low energies. However, this relation may undergo significant modifications at extremely high energy due to quantum gravitational effects, which can be expressed as~\cite{Acciari_2020}: 
\begin{equation}\label{?}
    E^2=p^2c^2\left[1-\sum_{n=1}^\infty \mathcal{S}\left(\frac{E}{E_{\text{QG},n}}\right)^n\right],
\end{equation}
where $\mathcal{S}=+1$ and $-1$ represent the subluminal and superluminal cases, respectively. Here, $E_{\text{QG},n}$ denotes the $n$-th order quantum gravity energy scale, which is typically associated with the Planck energy. 
This modification implies that the group velocity of photons is no longer a constant but instead becomes energy-dependent as: 
\begin{equation} \label{deltav}
    v_{\gamma}(E)=\frac{\partial E}{\partial p}\simeq c\left[1-\mathcal{S}\frac{n+1}{2}\left(\frac{E}{E_{\text{QG},n}}\right)^n\right].
\end{equation}
Given that higher-order effects can often be negligible, we restrict our analysis to the \ac{1st-order} ($n = 1$) and \ac{2nd-order} ($n = 2$) LIV scenarios.

To explore potential \ac{LIV} effects, we analyze high-energy photons emitted from a cosmological source and subsequently detected on Earth. Let $t_{i,n}^{\text{obs}}$ denote the observed arrival time of the $i$-th photon subjected to the $n$-th order LIV effect, while $t_{i,n}^{\text{i}}$ represents the corresponding arrival time under the assumption of Lorentz invariance.  
They are related through: 
\begin{equation}\label{???}
    t_{i,n}^{\text{i}}= t_{i,n}^{\text{obs}} - \mathcal{S}\frac{n+1}{2}\frac{E_i^n}{E^n_{\text{QG},n}}\int_0^{z_s}dz\frac{(1+z)^n}{H(z)} \equiv  t_{i,n}^{\text{obs}}-\tau_n E_i^n,
\end{equation}
where $z_s$ denotes the redshift of the source, $E_i$ represents the energy of the $i$-th photon, $\tau_n$ is a parameter that incorporates both $\mathcal{S}$ and $E_{\text{QG},n}$. The Hubble parameter $H(z)$ is given by $H(z)=H_0\sqrt{\Omega_m(1+z)^3+\Omega_\Lambda}$,
where $H_0= 67.36 \; \text{km}\ \text{s}^{-1} \text{Mpc}^{-1}$ is the Hubble constant, and $\Omega_m = 0.315$ and $\Omega_\Lambda = 0.685$ are the current matter and dark energy density parameters \cite{Planck_2018}, respectively. 
For practical calculations, we define the following dimensionless parameters: $\eta_1 = \mathcal{S} E_{\text{PL}} / E_{\text{QG},1}$ and $\eta_2 = 10^{-16} \cdot \mathcal{S} E^{2}_{\text{pl}} / E^2_{\text{QG},2}$, where $ E_{\text{pl}}$ denotes the Planck energy scale. 
For GRB 221009A with a redshift of $z = 0.151$, we derive $\tau_1 \cdot E = 5.865 \eta_1 \tilde{E}~\rm{s}$ and $\tau_2 \cdot E^2 = 7.768 \eta_2 \tilde{E}^2~\rm{s}$, where $\tilde{E}$ represents the photon energy in units of TeV. 

If LIV exits, the observed photon arrival time profile of $t_{i,n}^{\text{obs}}$ would include LIV effects, while the derived profile of $t_{i,n}^{\text{i}}$ would reflect the intrinsic temporal structure of the emission at the source \footnote{The intrinsic energy dependent emitting at the source is not considered.}. The appropriate LIV parameters should yield a physically plausible initial time profile, whereas incorrect parameters may introduce artificial structures in the reconstructed time profile \cite{Scargle2008}. 
For this analysis, we employ the DisScan algorithm \cite{Scargle2008}, designed for the analysis of time-energy tagged photon data, to investigate LIV effects and impose constraints. 
This algorithm operates by scanning the LIV parameter space to maximize the temporal sharpness of the reconstructed photon arrival time profile $t_{i,n}^{\text{i}}$, thereby recovering possible energy-dependent time delays.

Given photon data with time and energy measurements, along with a photon dispersion model, the algorithm primarily involves the following steps: 1. Construct the time profile of $t_{i,n}^{\text{i}}$ from the observed photon data. 
2. Define the Shannon entropy to quantify the sharpness of the reconstructed time profile. 
3. Determine the optimal LIV parameters by minimizing the Shannon entropy, corresponding to the physically meaningful time profile. 

In our analysis, we construct the time profile $t_{i,n}^{\text{i}}$ utilizing observations from \ac{LHAASO} of GRB 221009A. For a given LIV parameter $\tau_n$, we calculate $t_{i,n}^{\text{i}}$ using Eq. (\ref{???}). 
We consider a count-based profile of photons defined as: 
\begin{equation}\label{!}
    x_n = \frac{N_n}{\Delta t},
\end{equation}
where $N_n$ represents the photon count in the n-th time bin, and $\Delta t$ denotes the width of the time bin. Since higher energy photons are more sensitive to LIV parameters, we enhance their contribution by assigning them larger weights, thereby improving the overall sensitivity. Therefore, we also consider an energy-weighted profile defined as \cite{Scargle2008}

\begin{equation}\label{!!}
    x_n = \frac{\sum_{k=1}^{N_n} E^{(n)}_k}{\Delta t}, 
\end{equation}
where the summation runs over all photons in the $n$-th time bin, and $E^{(n)}_k$ denotes the energy of the $k$-th photon in this bin. 

To evaluate the quality of the derived profile, we employ Shannon entropy.
This concept, originally introduced by Claude Shannon in 1948 \cite{shannon1949mathematical}, serves as a measure of uncertainty within a system in the field of information theory and has broad applications across various disciplines. In physics, Shannon entropy is closely connected to thermodynamic entropy, as both concepts describe the degree of disorder or randomness within a system. The Shannon entropy is calculated as
\begin{equation}\label{SI}
    \text{I} = - \sum_n p_n \cdot \log p_n, 
\end{equation}
where $p_n \equiv x_n/\sum_n x_n$ denotes the normalized value of $x_n$. For a given $\tau$, we obtain the corresponding Shannon entropy. Generally, a high Shannon entropy indicates a random and featureless distribution, suggesting incorrect LIV values that significantly distort the intrinsic source profile \cite{Scargle2008}. Conversely, a low Shannon entropy implies a structured temporal profile, revealing optimal LIV parameters that effectively recover the initial emission characteristics. Thus, by minimizing the Shannon entropy across the LIV parameter space, we can determine the physically plausible LIV parameters that align with the observed data. 

\begin{figure}[htbp]
    \centering
    \begin{minipage}{0.32\textwidth}
        \centering
        \includegraphics[width=\linewidth]{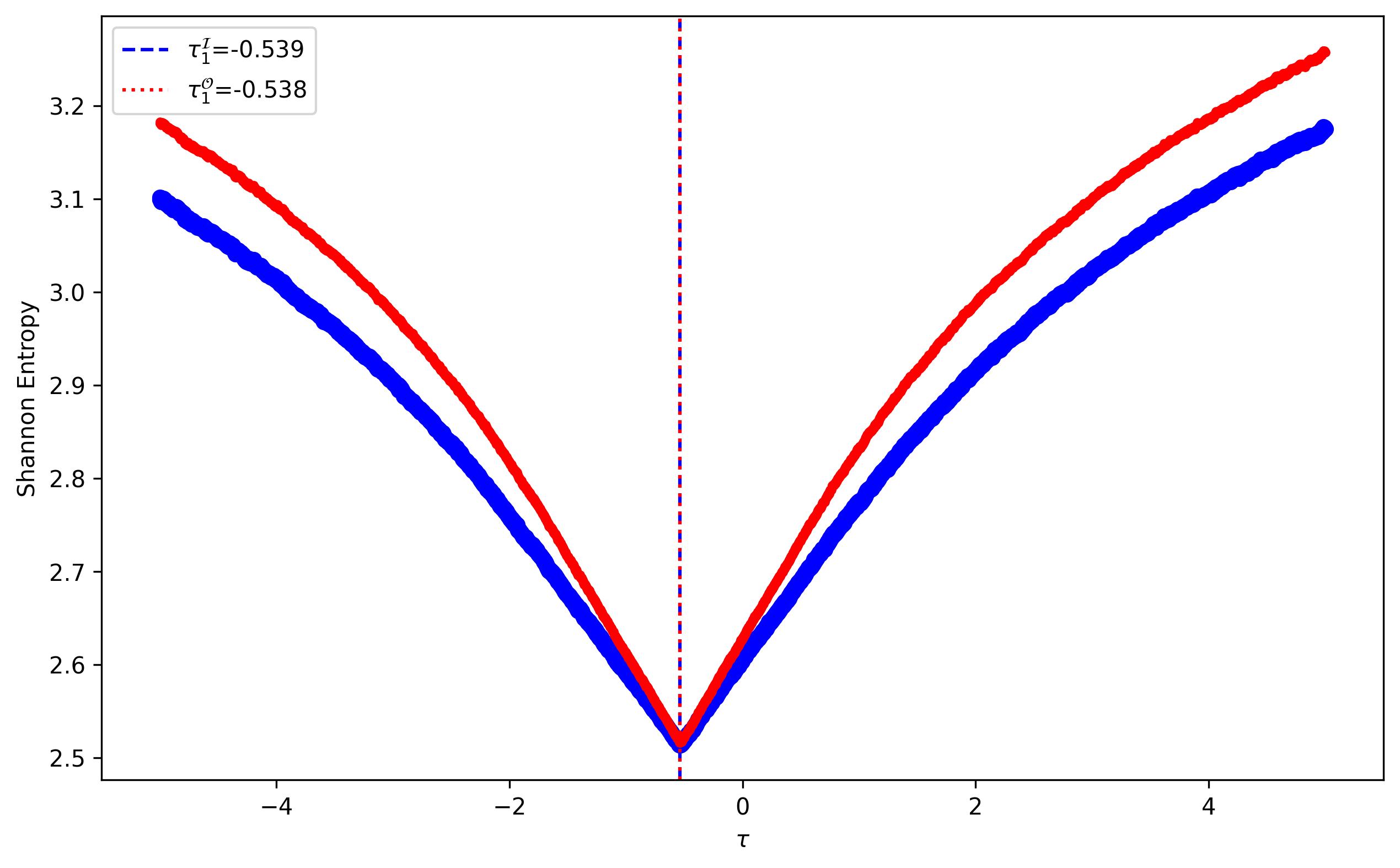} 
        %\caption{}
    \end{minipage}
    \hfill
    \begin{minipage}{0.32\textwidth}
        \centering
        \includegraphics[width=\linewidth]{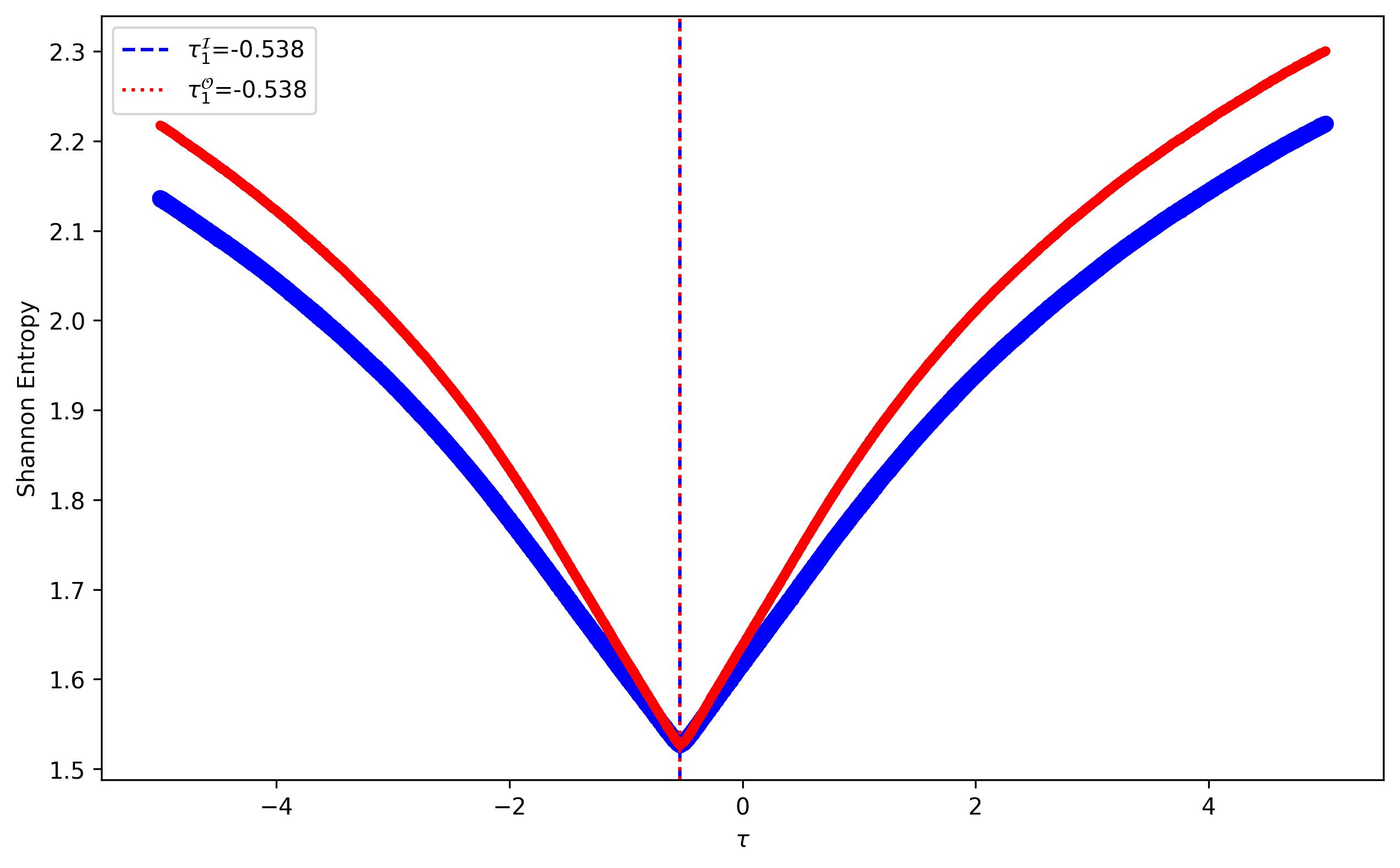} 
        %\caption{}
    \end{minipage}
    \hfill
    \begin{minipage}{0.32\textwidth}
        \centering
        \includegraphics[width=\linewidth]{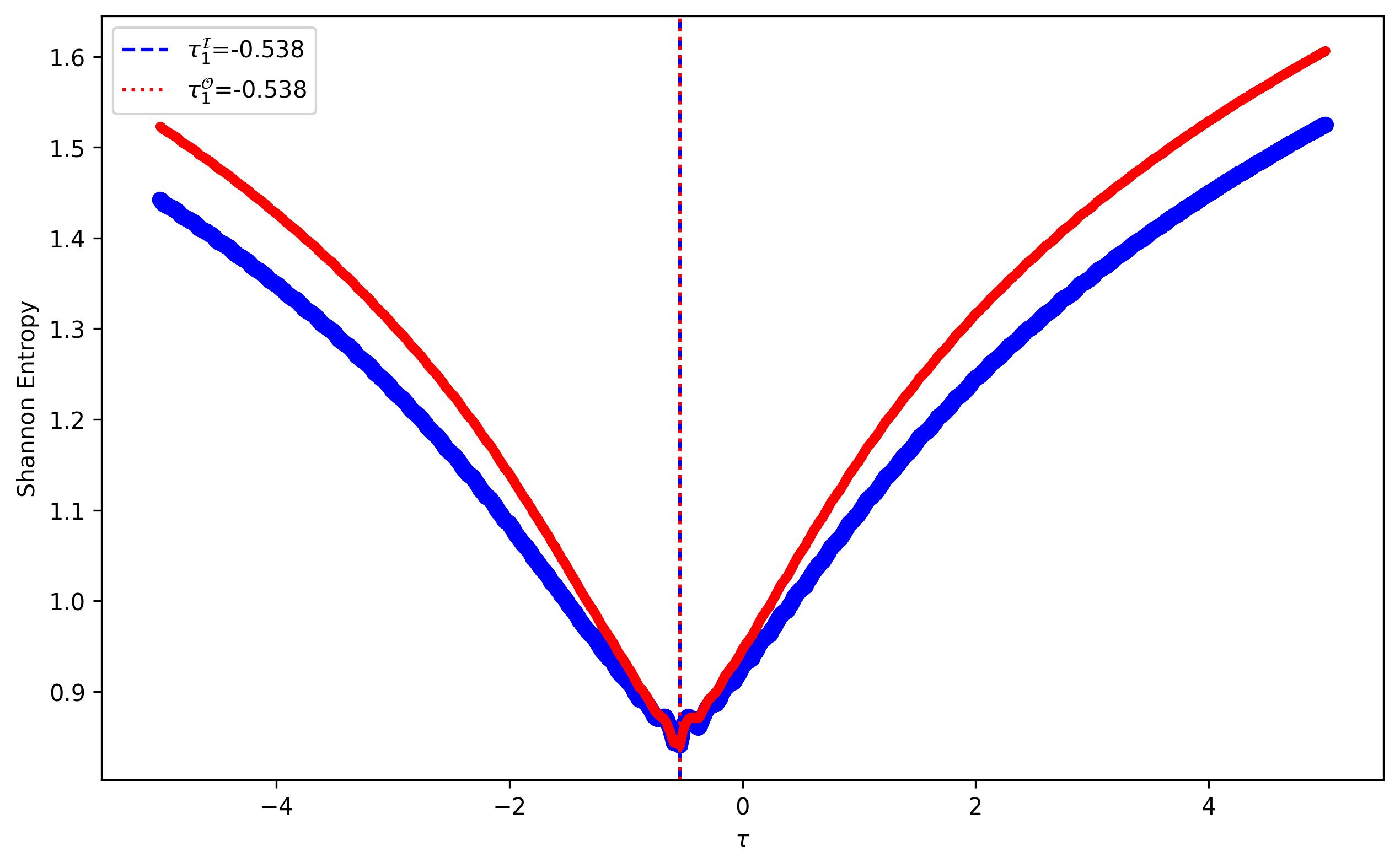} 
        %\caption{}
    \end{minipage}
    \caption{
    The distributions of Shannon entropy for the photons concerning the \ac{1st-order} \ac{LIV} parameter $\tau_1$ with time bin widths of 0.3~s, 3~s, and 15~s. 
    The photons are randomly generated over a uniform time interval from 0 to 100 s, incorporating the LIV effects with a parameter value of $\tau_1^* = -0.537$. The red and blue curves correspond to the results derived with the energy-weighted and photon-count profiles, respectively. The red and blue vertical dashed lines indicate the optimal $\tau_1$ values that minimize the Shannon entropy for the two cases, respectively.}
    \label{3-1}
\end{figure}

To demonstrate the efficacy of this methodology, we present three test cases employing simulated photon data. 
First, we generate a mock sample comprising $10^4$ photons distributed uniformly across a time interval from 0 to 100 seconds, incorporating LIV effects with a parameter value of $\tau_1^* = -0.537$. The energies of the photons are randomly generated within a uniform range of 0 to 100 TeV.  
Subsequently, we employ the DisCan algorithm to scan the LIV parameters with this data sample. The resulting distributions of Shannon entropy are displayed in Fig.~\ref{3-1} for three distinct time bin widths. 
The blue and red curves represent the results corresponding to the energy-weighted and count-based profiles, respectively. We observe that both profiles correctly identify the input LIV parameter through their respective entropy minima, thereby validating the algorithm's capability to reconstruct sharply distributed temporal profiles. 
Furthermore, the final results are not significantly concerned with time bin intervals in this case.  

\begin{figure}[htbp]
    \centering
    \begin{minipage}{0.32\textwidth}
        \centering
        \includegraphics[width=\linewidth]{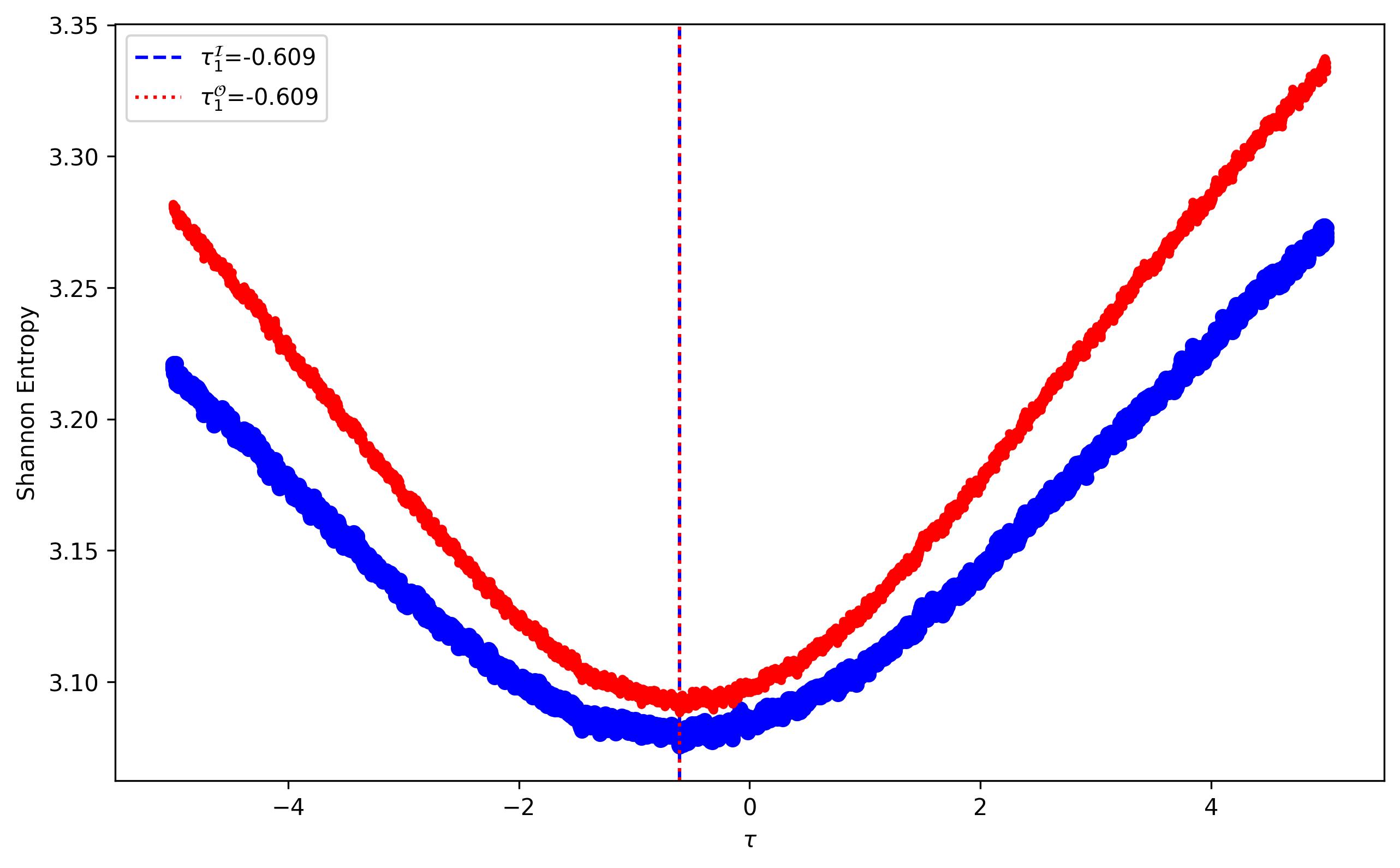}
        %\caption{}
    \end{minipage}
    \hfill
    \begin{minipage}{0.32\textwidth}
        \centering
        \includegraphics[width=\linewidth]{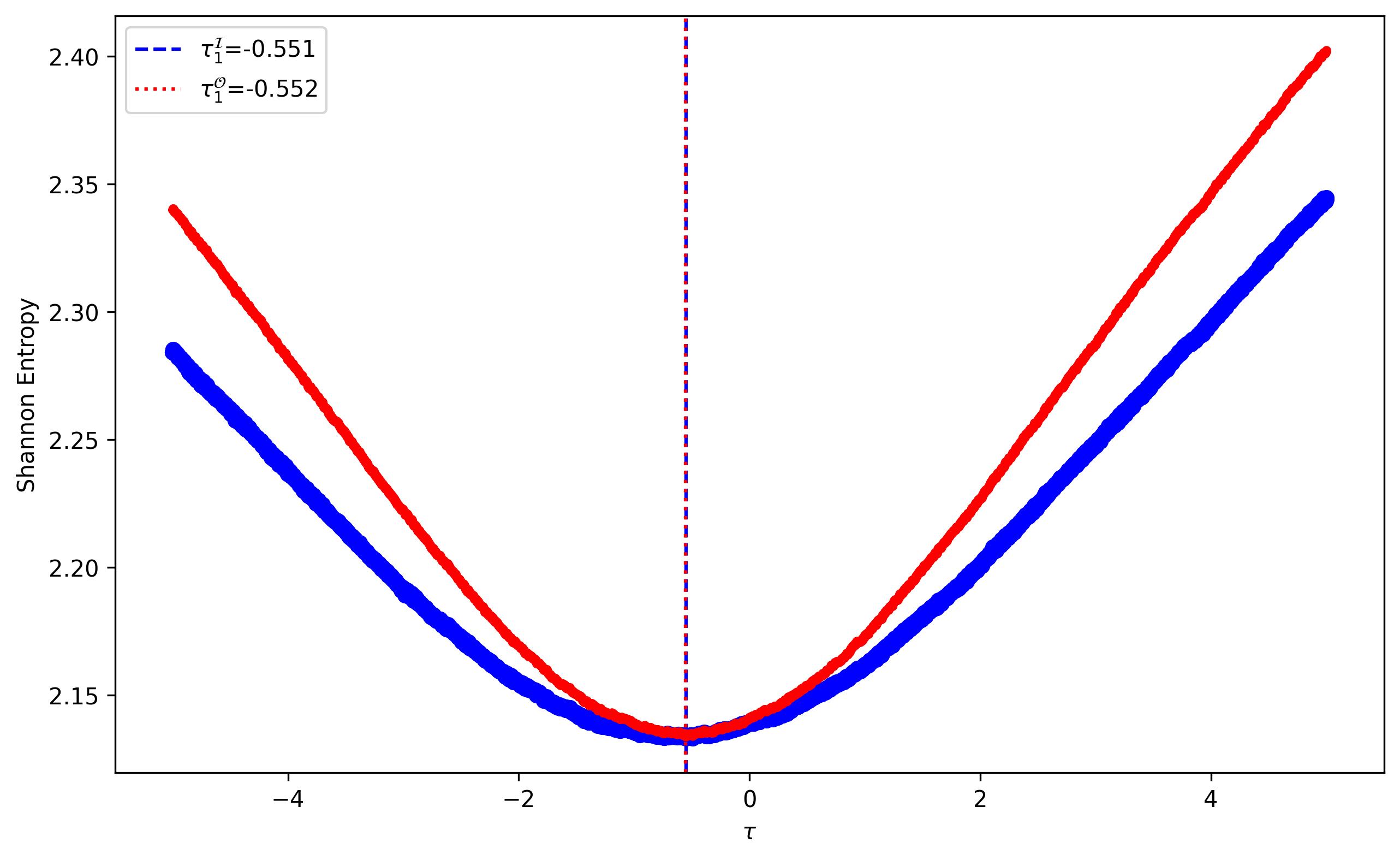}
        %\caption{}
    \end{minipage}
    \hfill
    \begin{minipage}{0.32\textwidth}
        \centering
        \includegraphics[width=\linewidth]{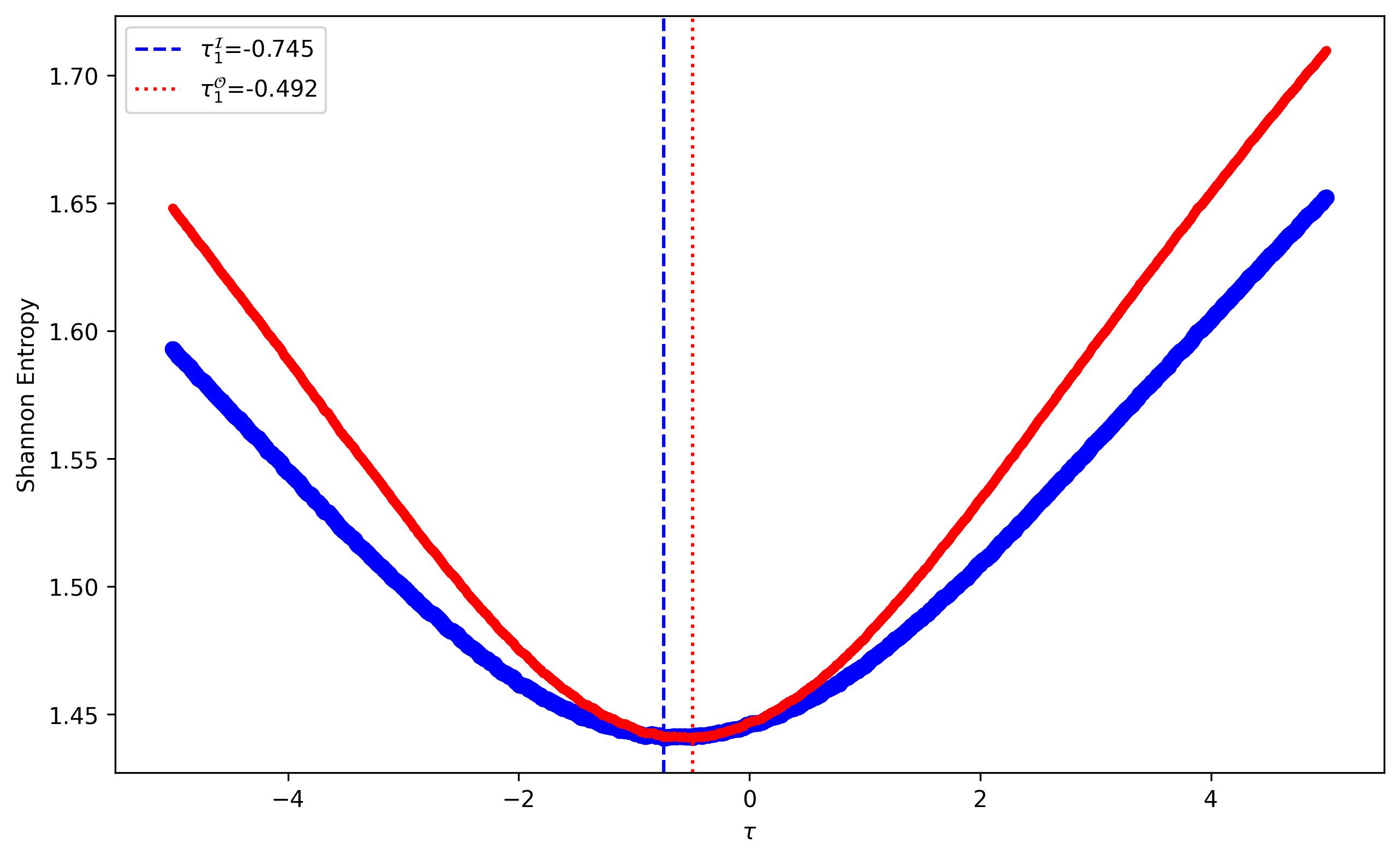}
        %\caption{}
    \end{minipage}
    \caption{
    The same as Fig.~\ref{3-1}, but for the arrival time profile of photons generated with a normal distribution.} 
    \label{3-2}
\end{figure}

We also generate another sample where the photon arrival times follow a normal distribution with a mean of zero and a standard deviation of 100 s, representing a more gradual temporal profile. The corresponding results are illustrated in Fig.~\ref{3-2}. 
We observe that the reconstruction in Fig.~\ref{3-2} is less precise compared to the previous case. In this case, the initial time profile has a less sharp edge, which diminishes the algorithm's ability to identify specific features crucial for the entropy minimization procedure. Nevertheless, the algorithm still converges toward the vicinity of the true LIV parameter, thereby demonstrating its validity. 

\begin{figure}[htbp]
    \centering
    \begin{minipage}{0.32\textwidth}
        \centering
        \includegraphics[width=\linewidth]{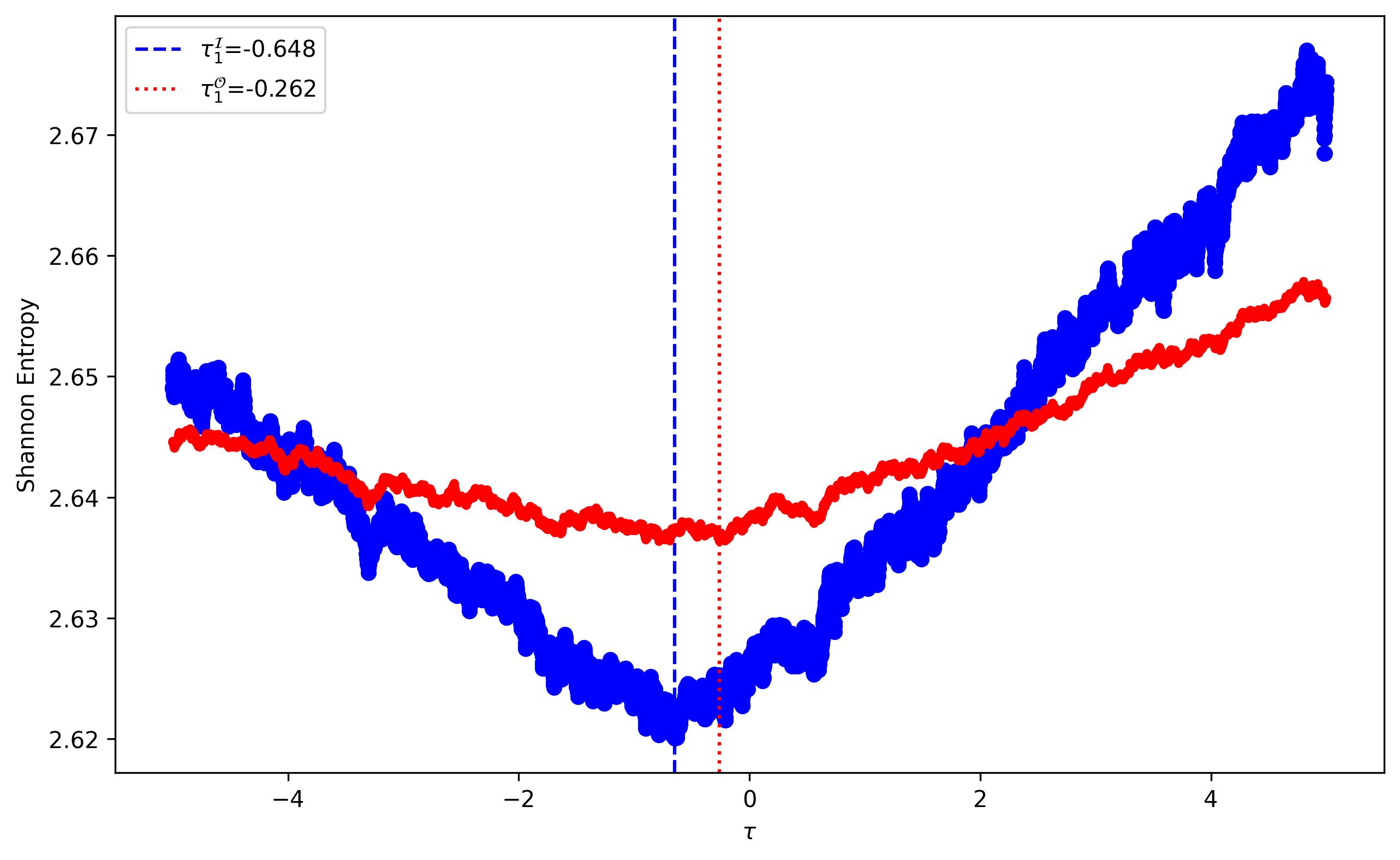}
        %\caption{}
    \end{minipage}
    \hfill
    \begin{minipage}{0.32\textwidth}
        \centering
        \includegraphics[width=\linewidth]{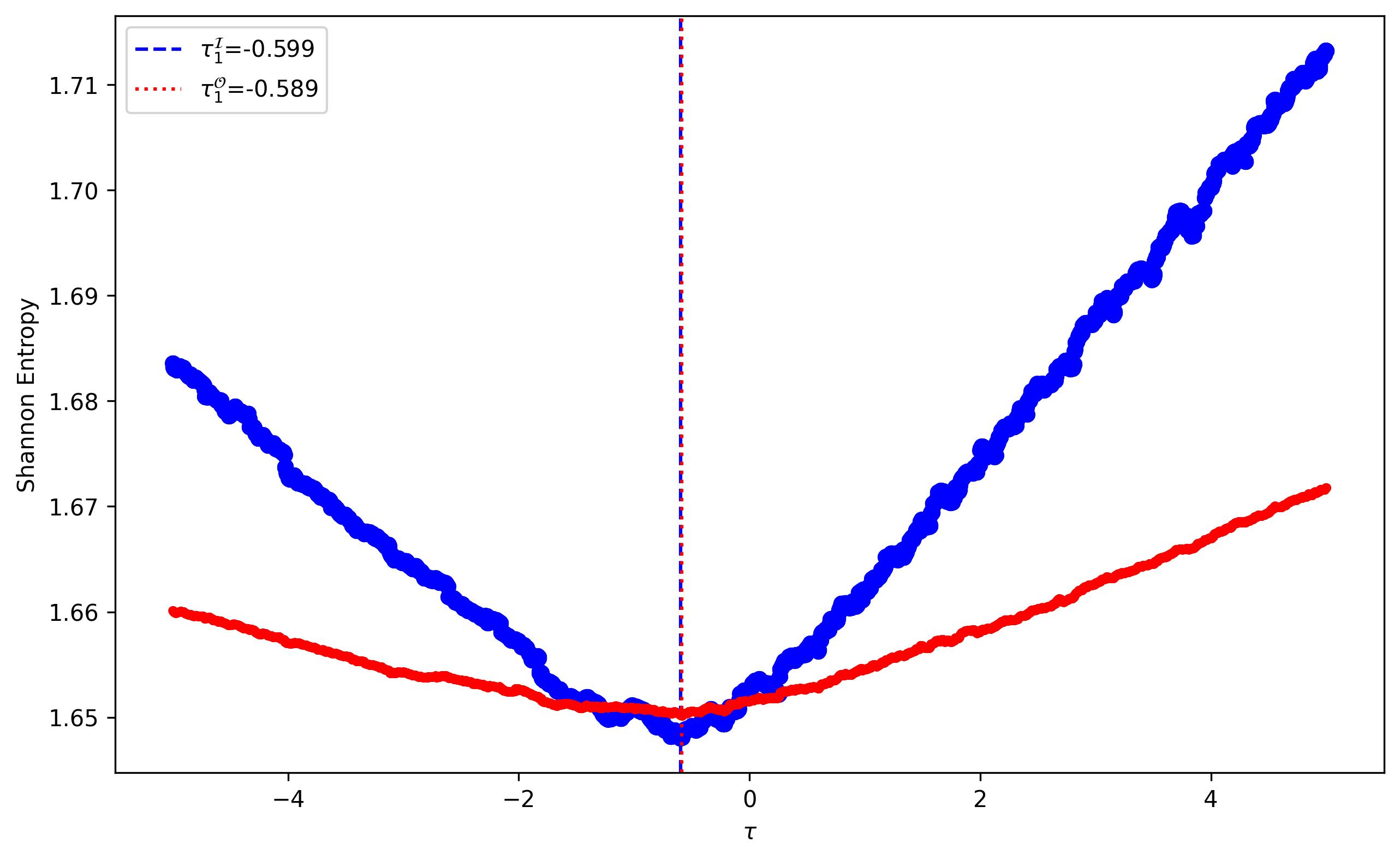}
        %\caption{}
    \end{minipage}
    \hfill
    \begin{minipage}{0.32\textwidth}
        \centering
        \includegraphics[width=\linewidth]{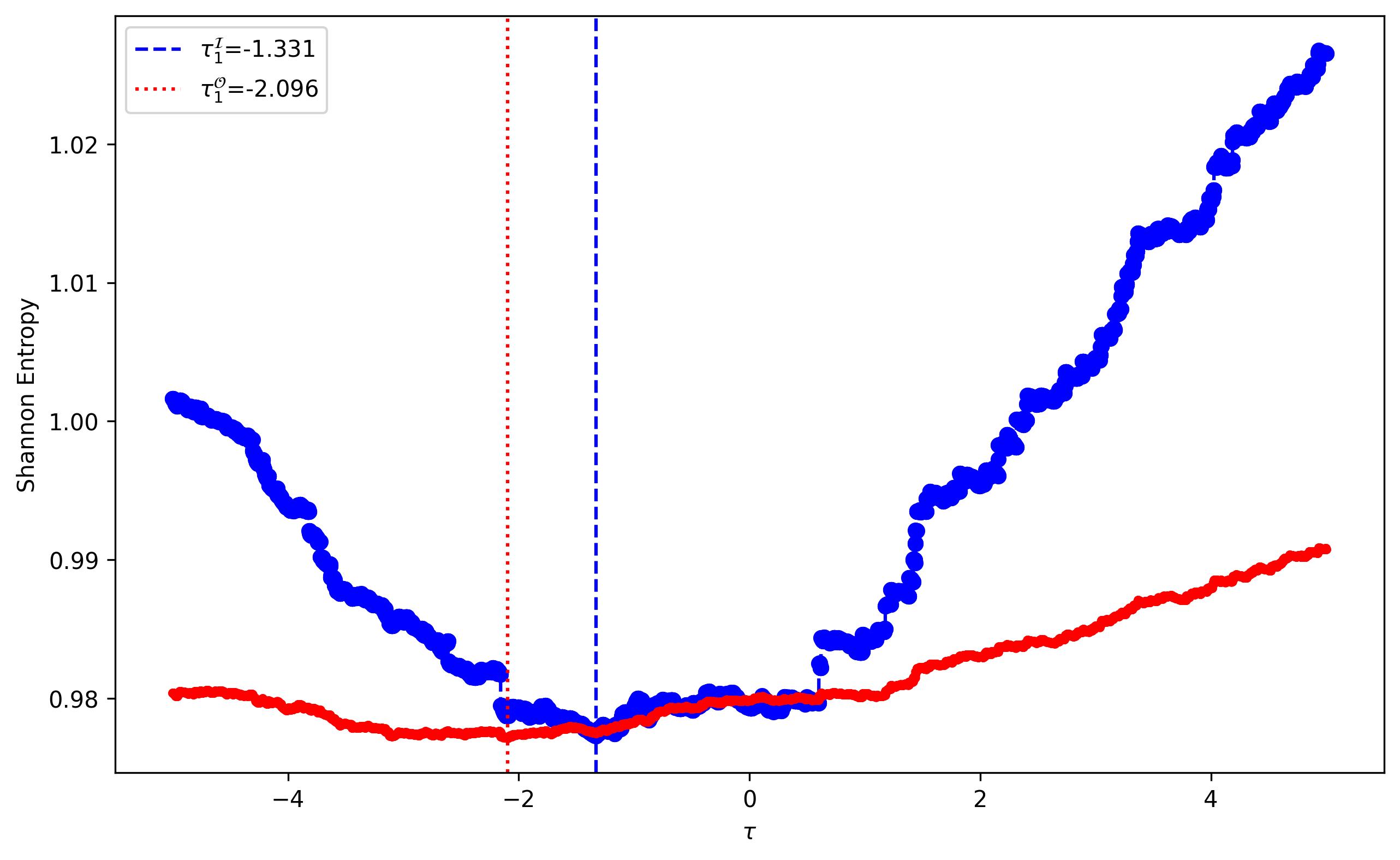}
        %\caption{}
    \end{minipage}
    \caption{
    The same as Fig.~\ref{3-1}, but for the photons generated using the time profile and energy spectrum reported by the LHAASO collaboration. 
    } 
    \label{3-3}
\end{figure}

To further illustrate the feasibility of this method for LHAASO photon observations, we generate a sample of $10^4$ photons based on the time profile and energy spectrum of GRB 221009A reported by LHAASO. 
The time profile $\lambda (t)$ is expressed as follows \cite{2023_1}: 
\begin{equation} \label{lamda}
    \lambda(t)=\left\{\begin{array}{ll}
        0 &,\; t<226\\
        \left[\left(\frac{t_{b,0}}{t_{b,1}}\right)^{-\omega_1\alpha_1}+\left(\frac{t_{b,0}}{t_{b,1}}\right)^{-\omega_1\alpha_2}\right]^{-1/\omega_1}\left(\frac{t-226}{t_{b,0}}\right)^{\alpha_0}&,\; 226\leq t< t_{b,0}+226\\
        \left[\left(\frac{t-226}{t_{b,1}}\right)^{-\omega_1\alpha_1}+\left(\frac{t-226}{t_{b,1}}\right)^{-\omega_1\alpha_2}\right]^{-1/\omega_1}&,\; t_{b,0}+226\leq t< t_{b,2}+226\\
        \left[\left(\frac{t_{b,2}}{t_{b,1}}\right)^{-\omega_1\alpha_1}+\left(\frac{t_{b,2}}{t_{b,1}}\right)^{-\omega_1\alpha_2}\right]^{-1/\omega_1}\left(\frac{t-226}{t_{b,2}}\right)^{\alpha_3}&,\; t\geq t_{b,2}+226,\\
    \end{array}\right. 
\end{equation}
where the parameters $\left\{ \omega_1,\alpha_0, \alpha_1, \alpha_2,\alpha_3, t_{b,0}, t_{b,1}, t_{b,2} \right\}$ are chosen as $\left\{ 1.46,12.8,1.7,-1.108,\right.$ $\left.-1.87,4.85,16.1,560 \right\}$, respectively. The energy spectrum is adopted as a power-law form as $dN(E) / dE = E^{-2.579}$. 
We employ rejection sampling and inverse transform sampling to generate photon arrival times within [230, 400] sand energies within [0.3, 10] TeV, respectively. 
Fig.~\ref{3-3} shows the resulting distributions of Shannon entropy for time bin widths of 0.3 s, 3 s and 15 s. We can see that the optimal value of $\tau$ aligns well with the input $\tau^*$ for the time bin width of 3 s.

\begin{figure}[htbp]
    \centering
    \begin{minipage}{0.32\textwidth}
        \centering
        \includegraphics[width=\linewidth]{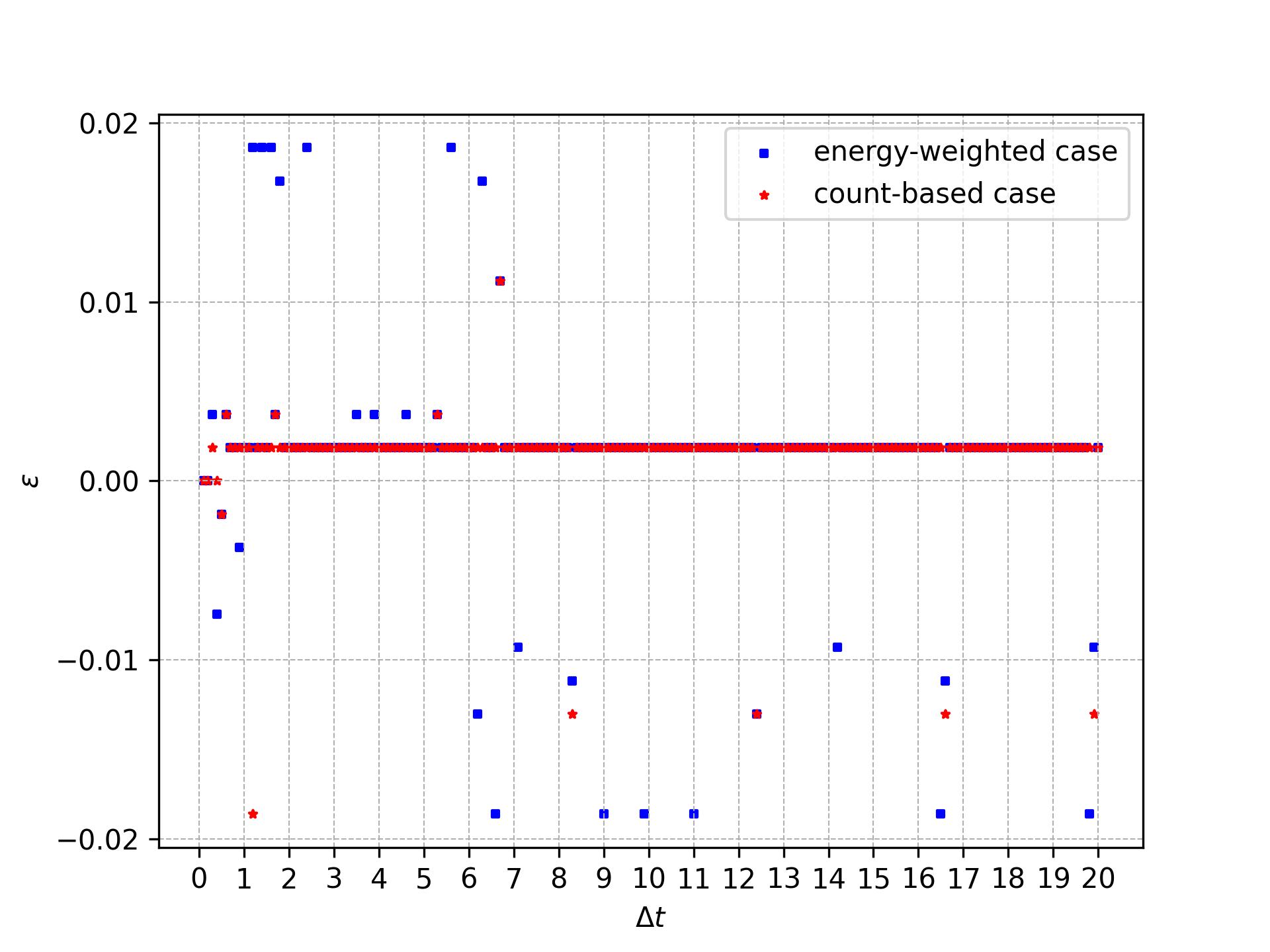}
        %\caption{}
    \end{minipage}
    \hfill
    \begin{minipage}{0.32\textwidth}
        \centering
        \includegraphics[width=\linewidth]{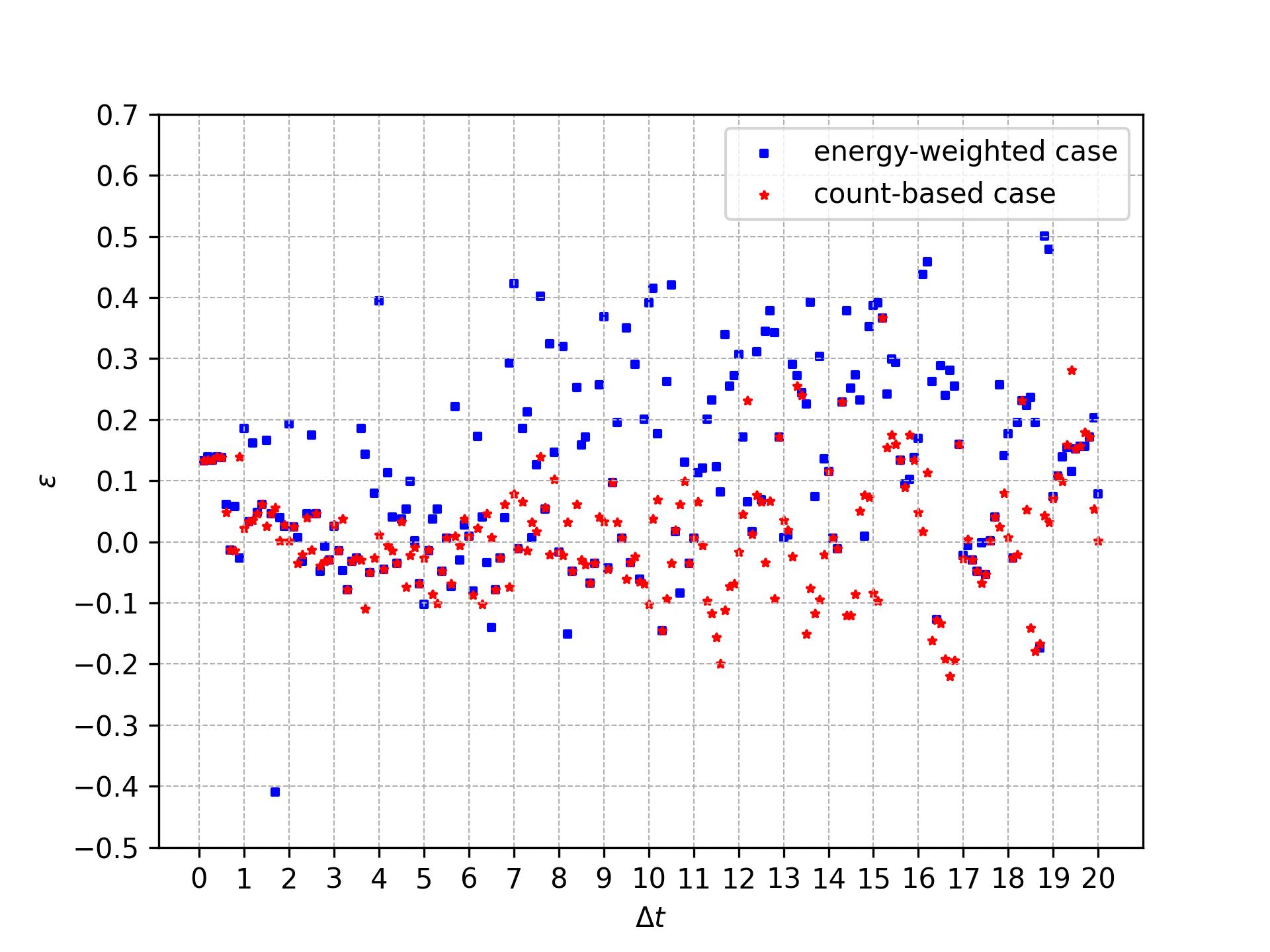}
        %\caption{}
    \end{minipage}
    \hfill
    \begin{minipage}{0.32\textwidth}
        \centering
        \includegraphics[width=\linewidth]{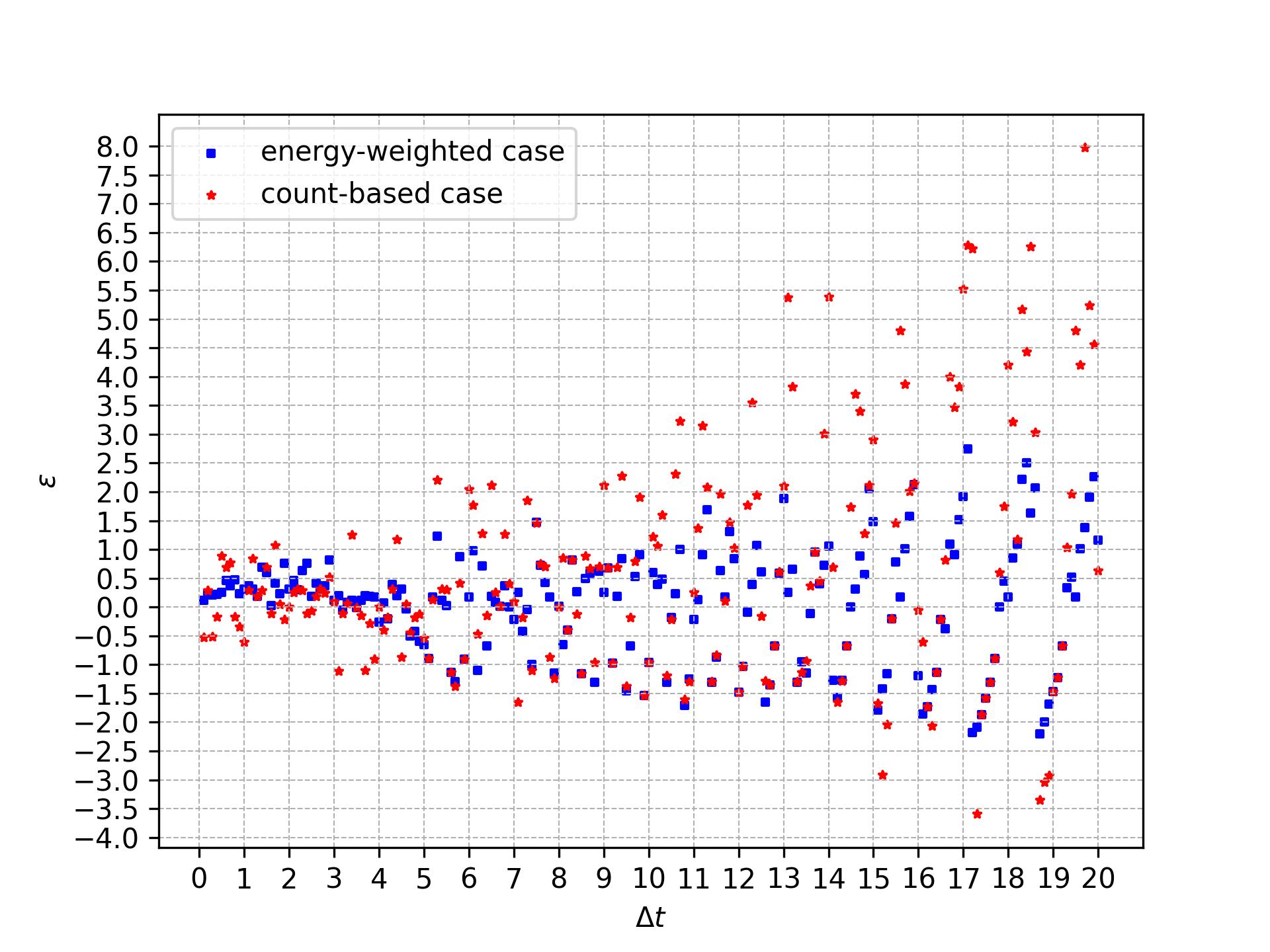}
        %\caption{}
    \end{minipage}
    \caption{
    The dependence of the optimal parameter $\tau$ on $\Delta t$. The left, middle and right panels correspond to the cases shown in Fig.~\ref{3-1}, Fig.~\ref{3-2} and Fig.~\ref{3-3}, respectively. 
    } 
    \label{3-4}
\end{figure}

We quantify the reconstruction accuracy using the relative error $\epsilon \equiv \left(\tau^* - \bar{\tau}\right) / \left| \tau^* \right|$, where $\bar{\tau}$ denotes the optimal value minimizing Shannon entropy. Fig.~\ref{3-4} presents the dependence of $\epsilon$ on the time bin width $\Delta t$ for the three generated samples. 
Our results indicate that for the sample generated with the reported time profile from LHAASO, the energy-weighted estimator for $\bar{\tau}$ yields more precise results compared to the count-based estimator. Furthermore, smaller time bin widths provide better results. 

\section{Exploring LIV effects using the observational data of GRB221009A from LHAASO}\label{sec:mine}

In order to investigate the potential \ac{LIV} effects, we utilize the LHAASO data of GRB 221009A, collected from both the \ac{KM2A} and \ac{WCDA} detectors. 
The \ac{KM2A} detector provides high-precision measurements of the arrival time and energy of each photon \cite{2023_2}. In contrast,  the \ac{WCDA} dataset features a coarser photon resolution \cite{2023_1}. Photons detected by the \ac{WCDA} detector are classified according to their $N_{\text{hit}}$ values, which correspond to the energy and size of secondary particle showers generated by high-energy photon interactions in the atmosphere \cite{2023_1}. Consequently, $N_{\text{hit}}$ serves as a proxy for estimating the energy of the original photon. 
The WCDA $N_{\text{hit}}$ values are grouped into the following intervals: [30, 33], [33, 40], [40, 63], [63, 100], [100, 250], and [250, $+ \infty$) \cite{2023_1}. For simplicity, we assume that photons within a given $N_{\text{hit}}$ interval share the same energy. For instance, photons in the [30, 33] interval are assigned an energy of 0.35 TeV~\cite{2023_1}. 
Furthermore, the WCDA timing data are recorded in 0.1-second bins, with each bin containing the corresponding photon count rate. Although this temporal resolution is relatively coarse compared to the KM2A data, it remains sufficient for our analysis. 

As indicated by Eq.~(\ref{deltav}), a greater disparity in photon energies leads to more pronounced differences in photon velocities, consequently inducing more significant distortions in the temporal profile of the photon signal. To maximize sensitivity to potential \ac{LIV} effects, our study employs both the \ac{KM2A} and \ac{WCDA} datasets, which collectively span a wide energy range.
The \ac{KM2A} detector provides high-precision energy measurements across an energy range of approximately 1-10 TeV, covering an order of magnitude higher than that of the \ac{WCDA}. We restrict our analysis to the \ac{WCDA} photons within the $N_{\text{hit}}$ bin of [30, 33], corresponding to the lowest energy photons. 

\begin{figure}[htbp]
    \centering
    \begin{minipage}{0.48\textwidth}
        \centering
        \includegraphics[width=\linewidth]{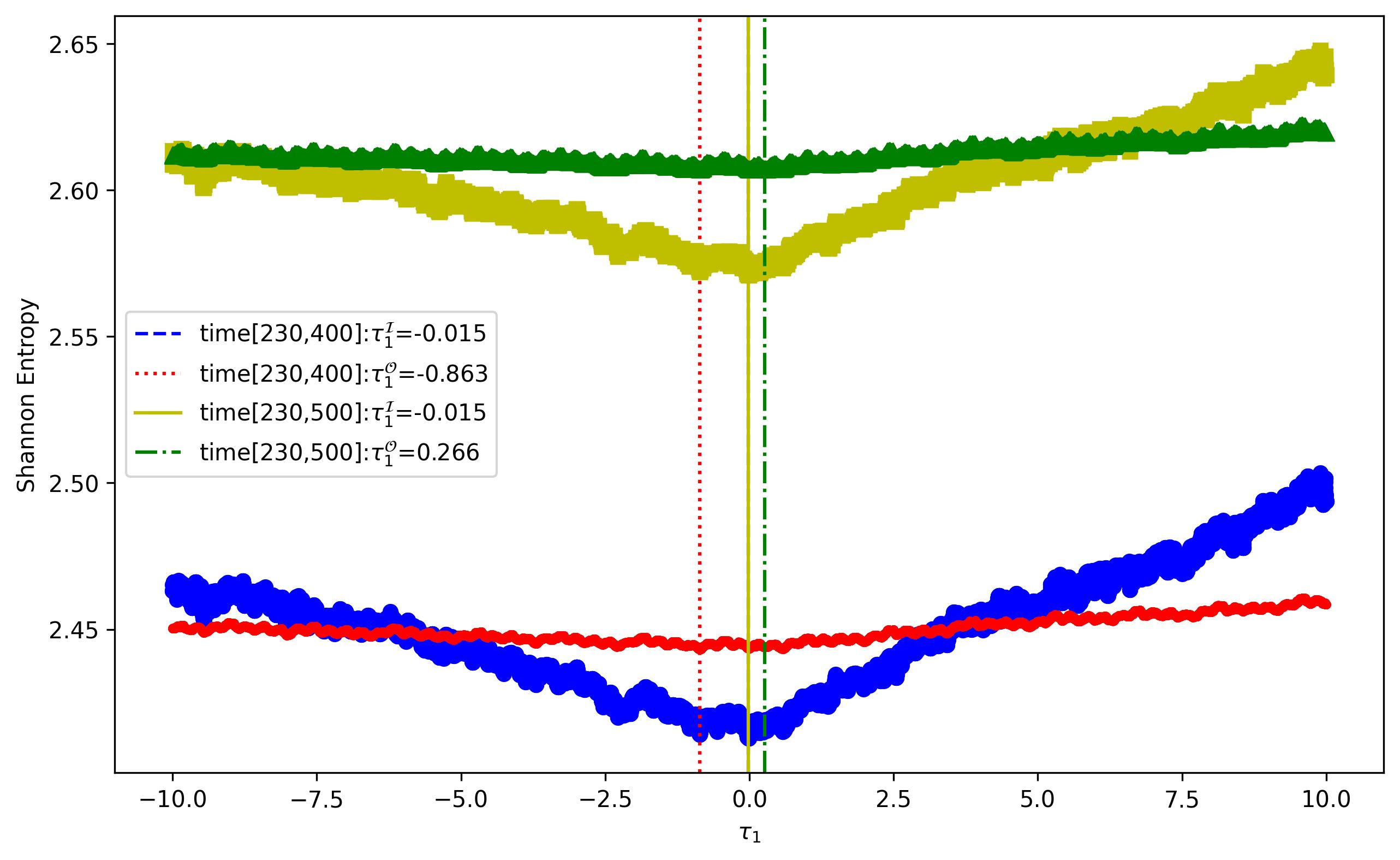} 
        %\caption{The graph depicting the variation of Shannon entropy with respect to the \ac{1st-order} \ac{LIV} parameter $\tau_1$ for a time bin interval of 2 second.} 
    \end{minipage}
    \hfill
    \begin{minipage}{0.48\textwidth}
        \centering
        \includegraphics[width=\linewidth]{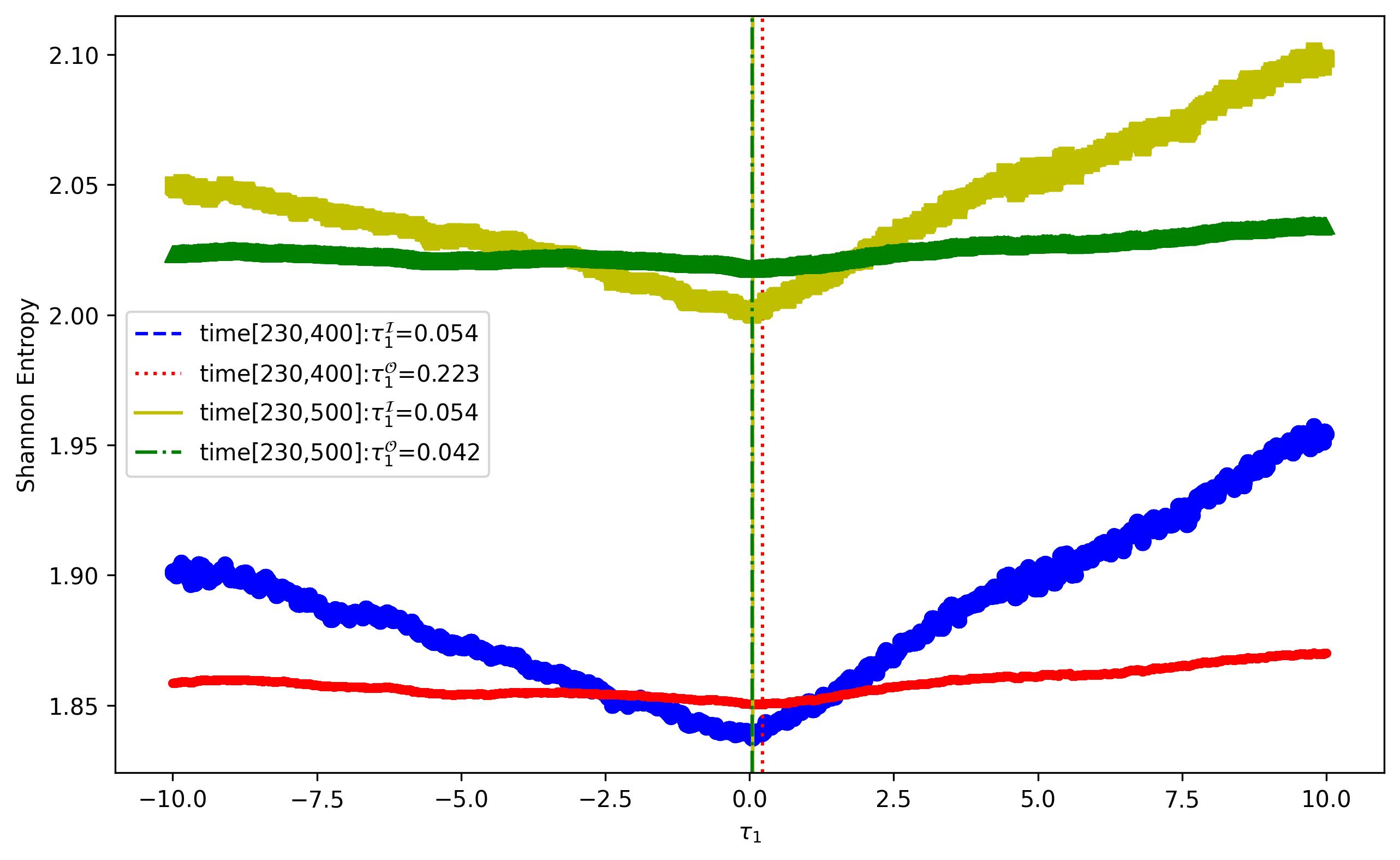} 
        %\caption{}
    \end{minipage}
    \hfill
    \begin{minipage}{0.48\textwidth}
        \centering
        \includegraphics[width=\linewidth]{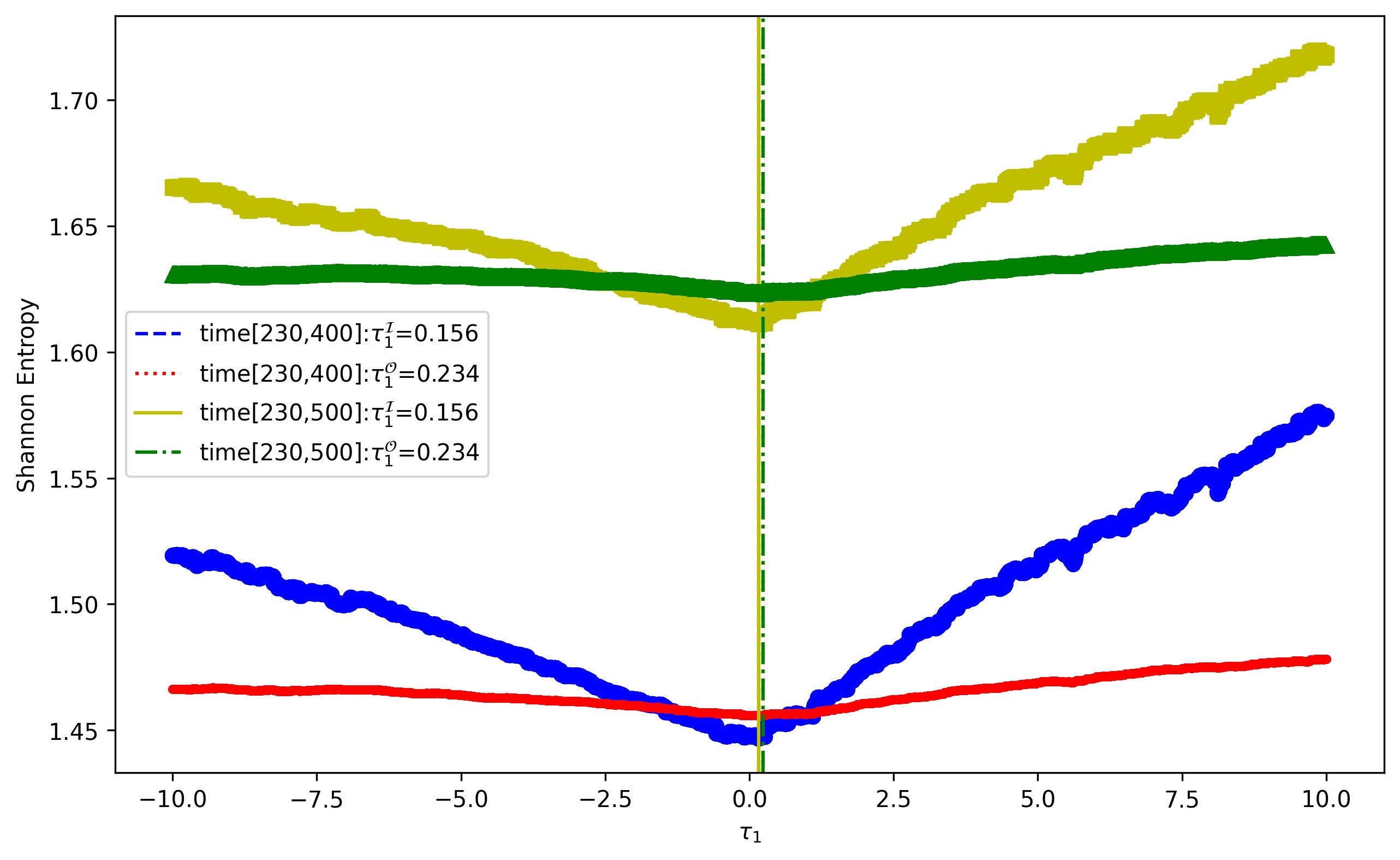} 
        %\caption{}
    \end{minipage}
    \hfill
    \begin{minipage}{0.48\textwidth}
        \centering
        \includegraphics[width=\linewidth]{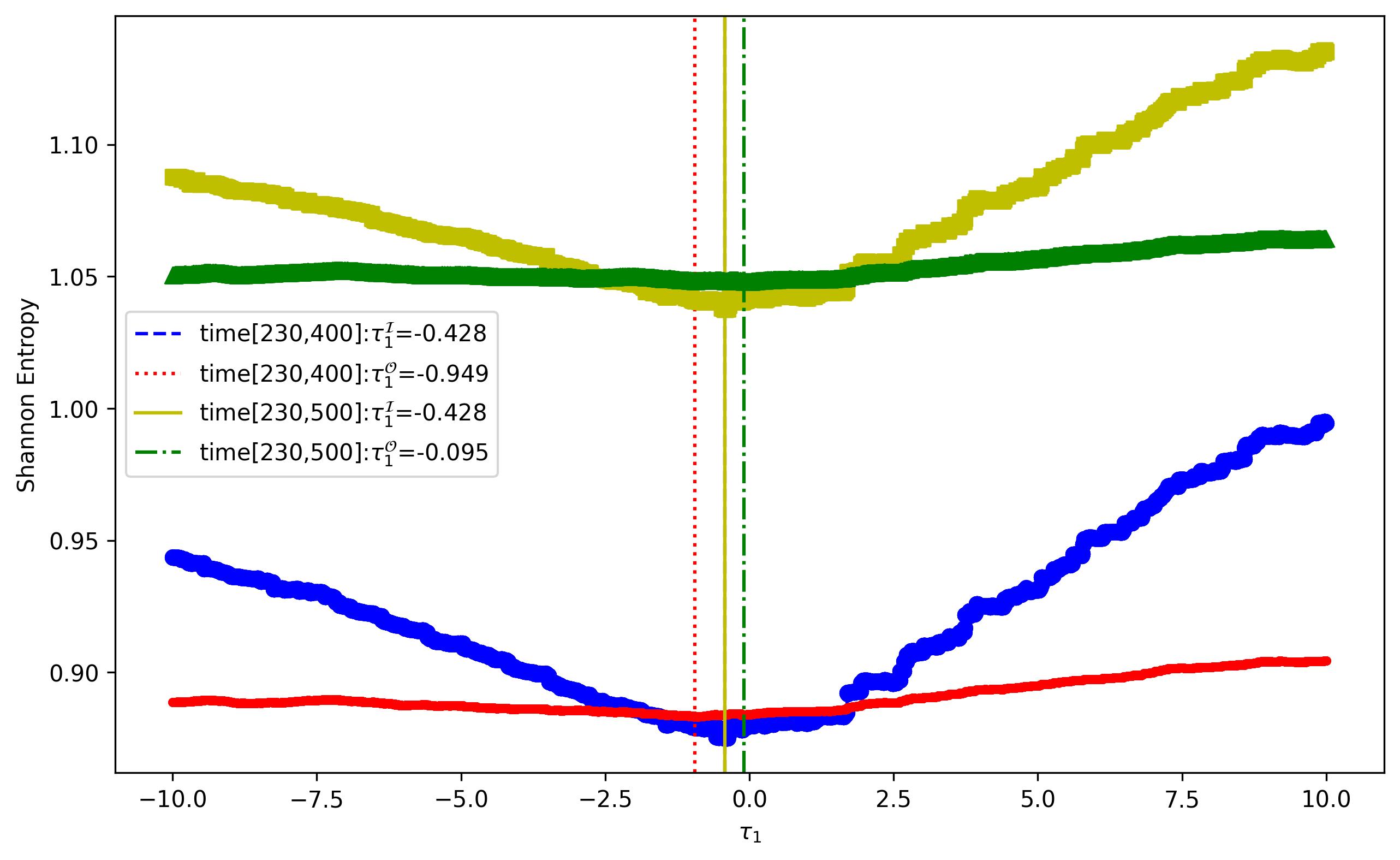} 
        %\caption{}
    \end{minipage}
    \caption{ 
    The variation of Shannon entropy with the \ac{1st-order} \ac{LIV} parameters $\tau_1$ for time binning intervals of 0.5~s, 2~s, 5~s, and 20~s, respectively. The red (blue) and green (yellow) curves correspond to the results in the [230,400] s and [230,500] s intervals analyzed using the count-based (energy-weighted) profile, respectively.
    }
    \label{4-1}
\end{figure}

\begin{figure}[htbp]
    \centering
    \begin{minipage}{0.48\textwidth}
        \centering
        \includegraphics[width=\linewidth]{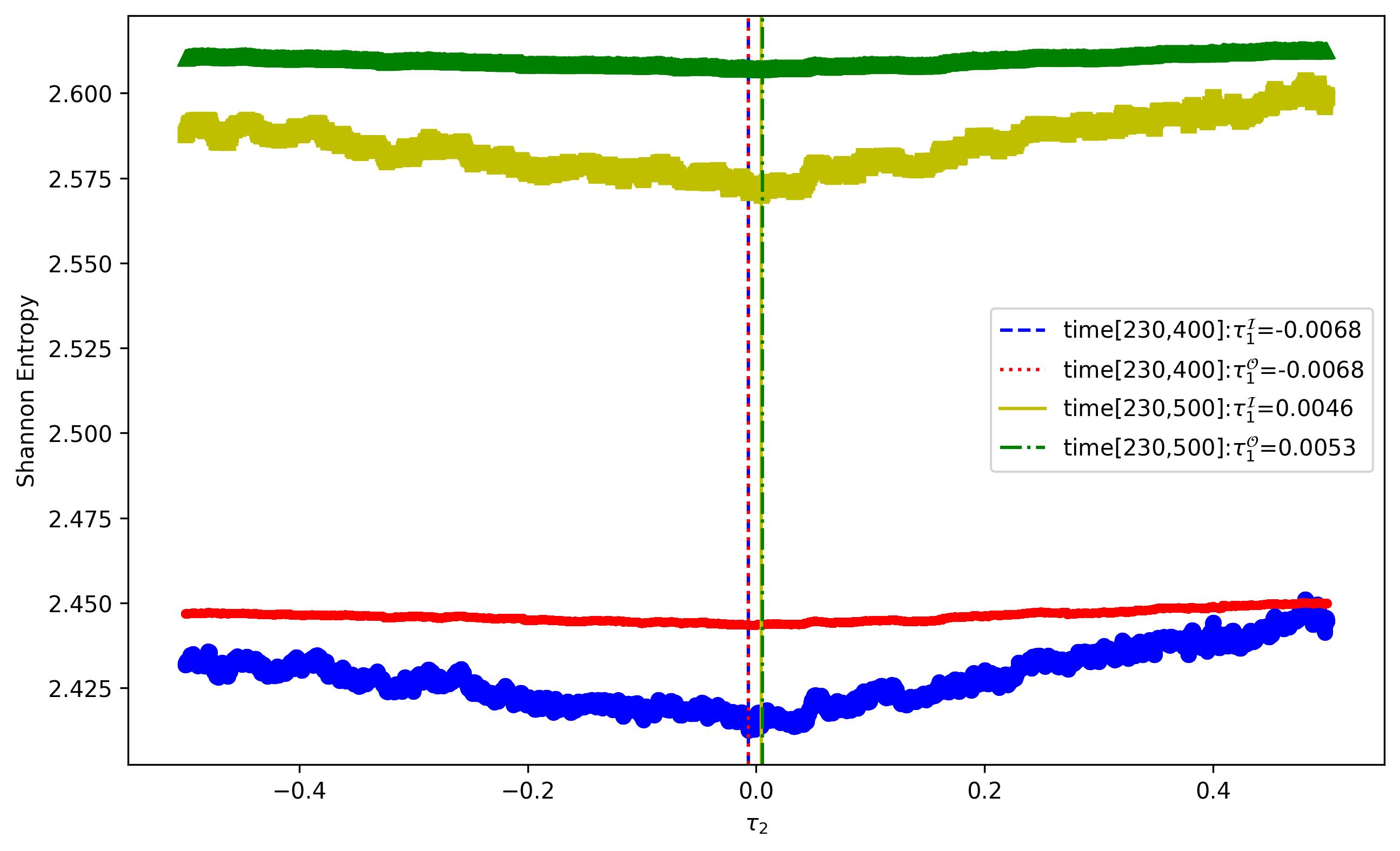} 
        %\caption{}
    \end{minipage}
    \hfill
    \begin{minipage}{0.48\textwidth}
        \centering
        \includegraphics[width=\linewidth]{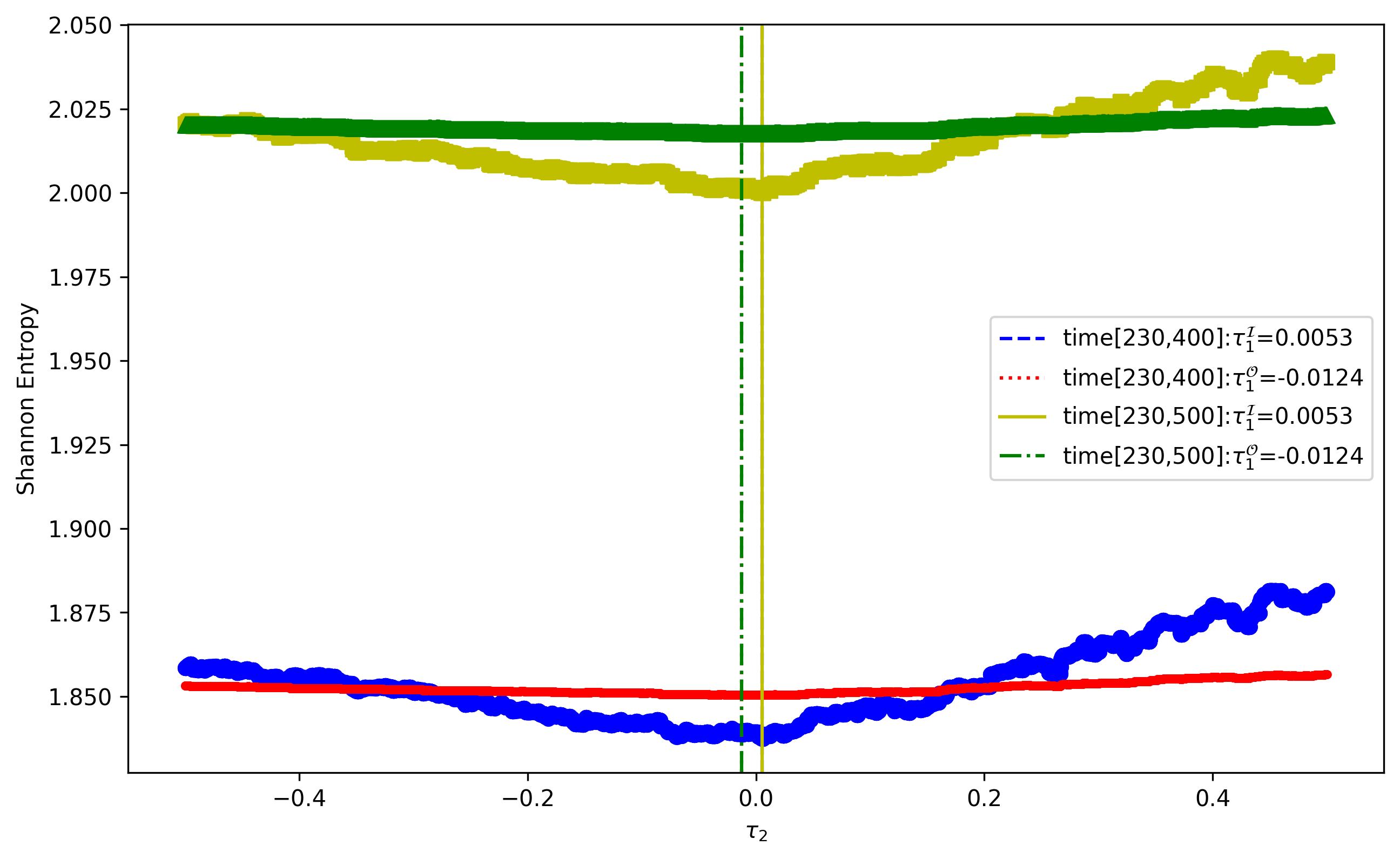} 
        %\caption{}
    \end{minipage}
    \hfill
    \begin{minipage}{0.48\textwidth}
        \centering
        \includegraphics[width=\linewidth]{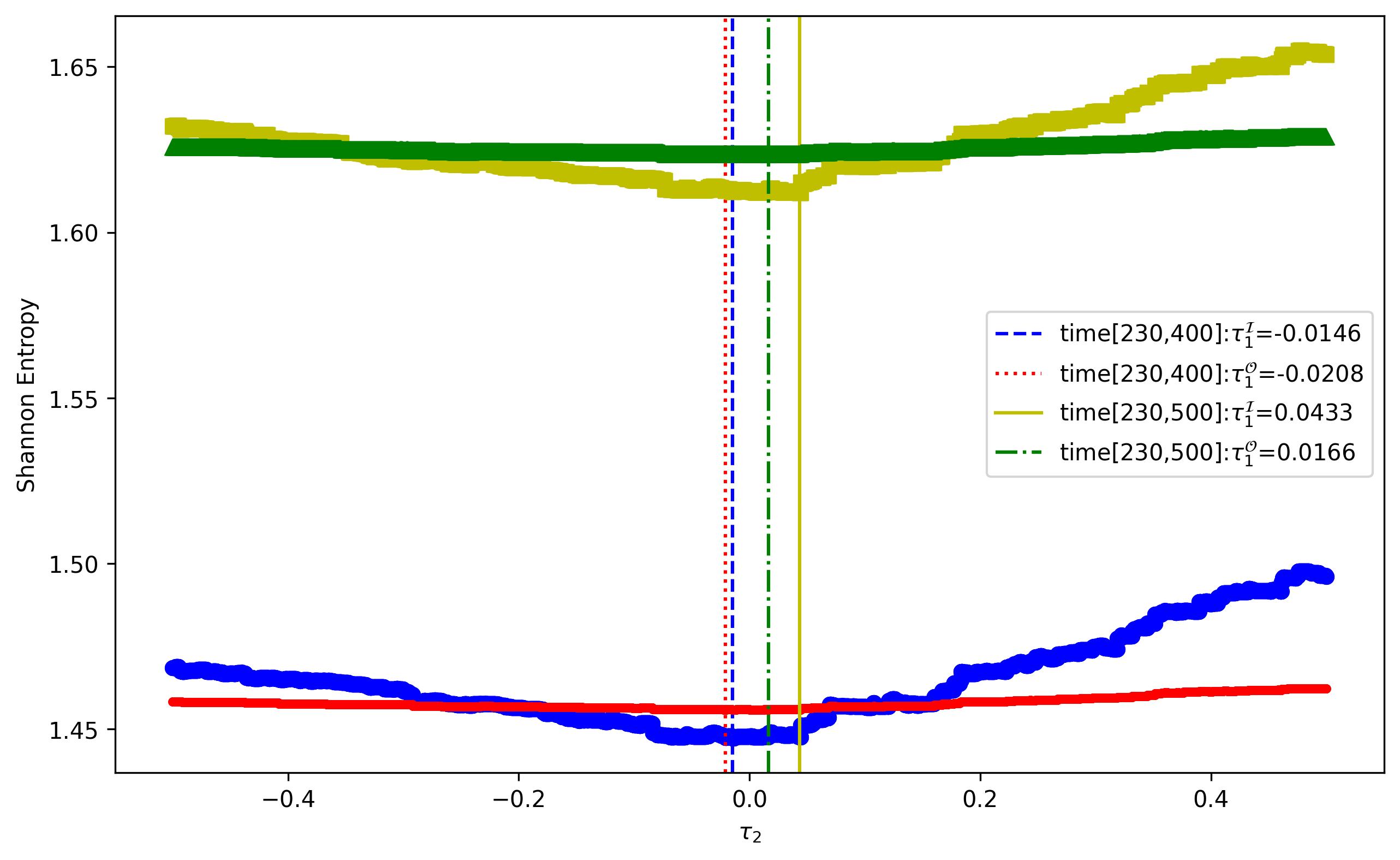} 
        %\caption{}
    \end{minipage}
    \hfill
    \begin{minipage}{0.48\textwidth}
        \centering
        \includegraphics[width=\linewidth]{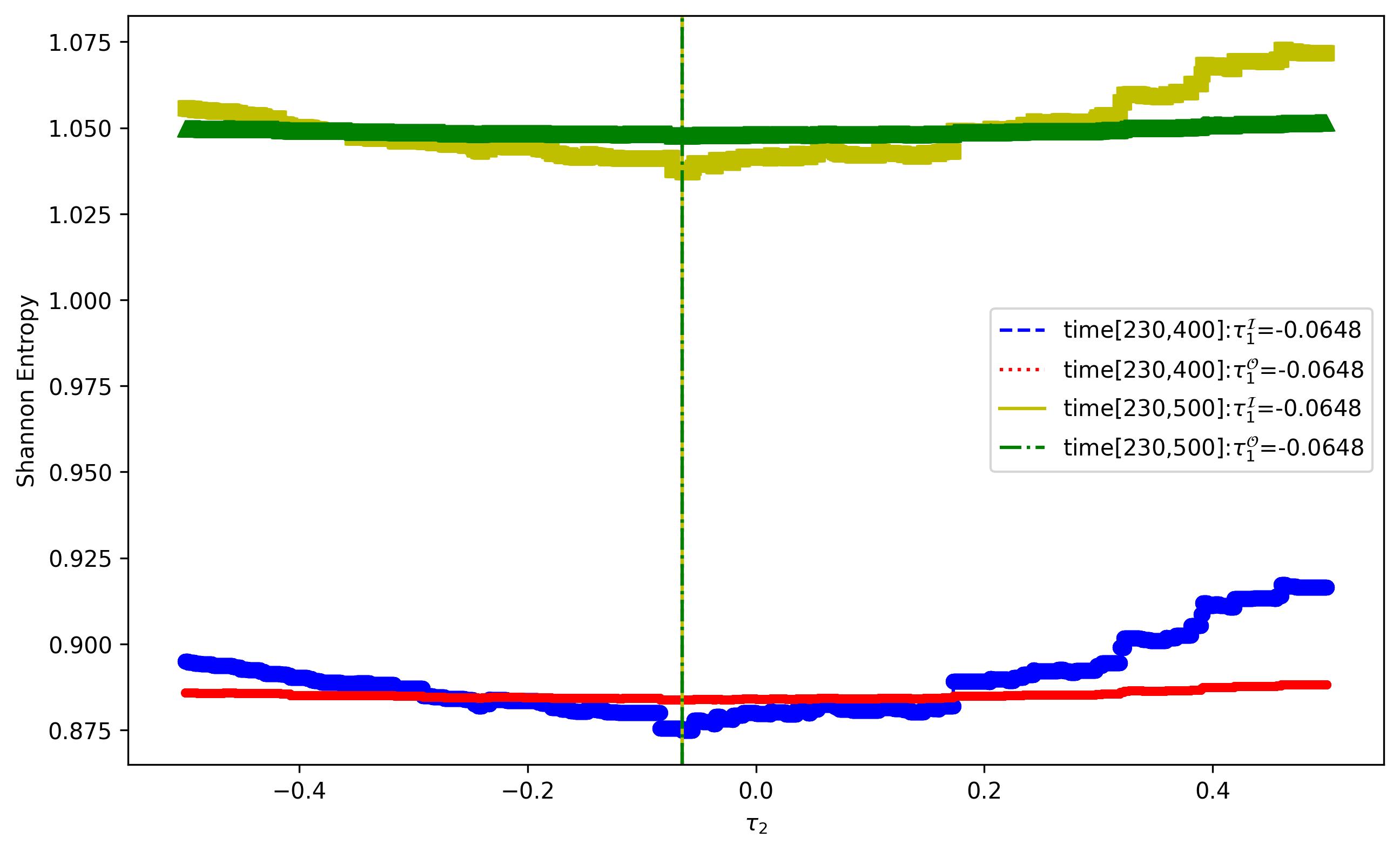} 
        %\caption{}
    \end{minipage}
    \caption{
    The same as Fig.~\ref{4-1}, but for the \ac{2nd-order} \ac{LIV} parameter $\tau_2$. 
    }
    \label{4-2}
\end{figure}

\begin{figure}[htbp]
    \centering
    \begin{minipage}{0.48\textwidth}
        \centering
        \includegraphics[width=\linewidth]{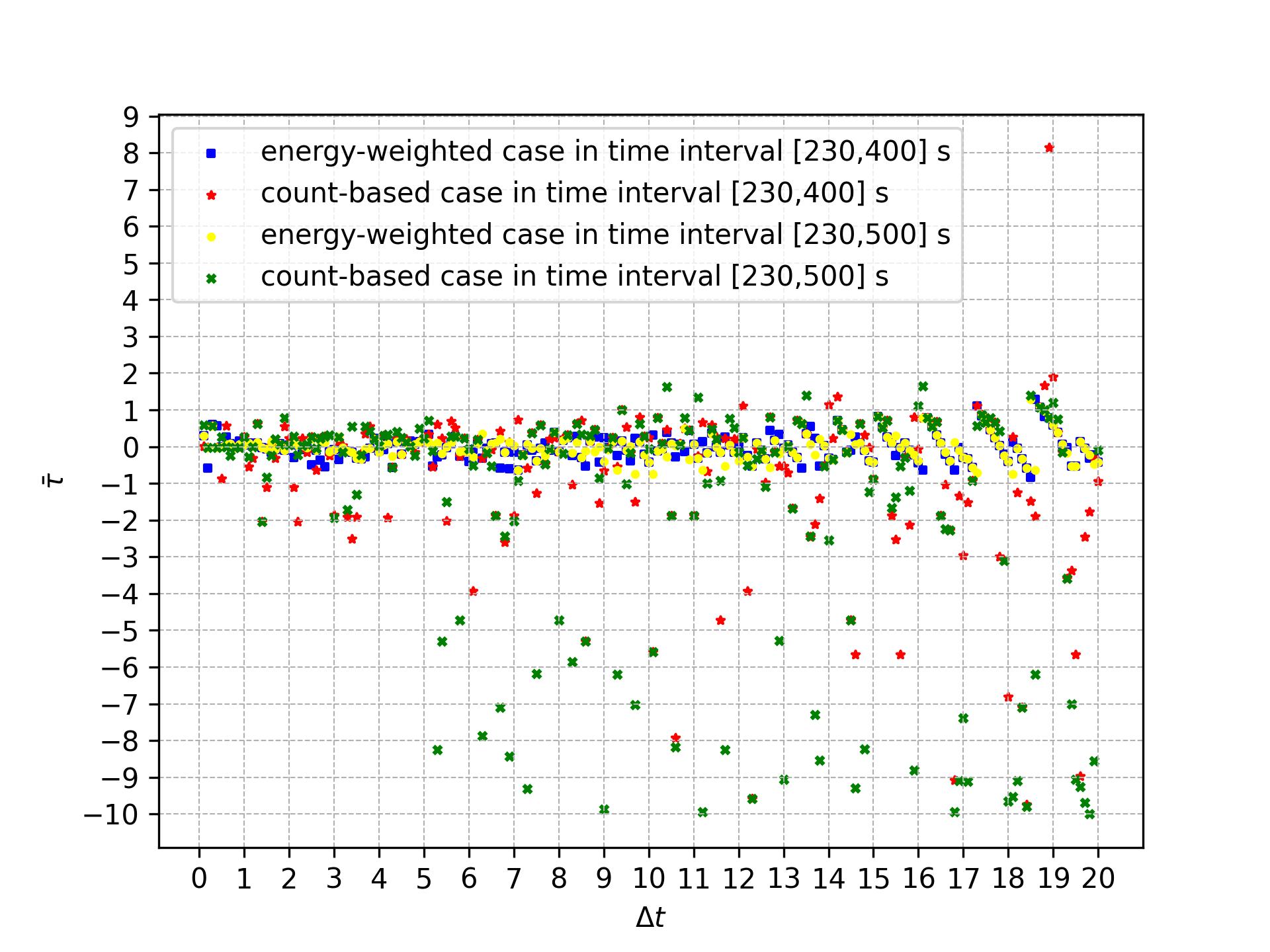} 
        %\caption{}
    \end{minipage}
    \hfill
    \begin{minipage}{0.48\textwidth}
        \centering
        \includegraphics[width=\linewidth]{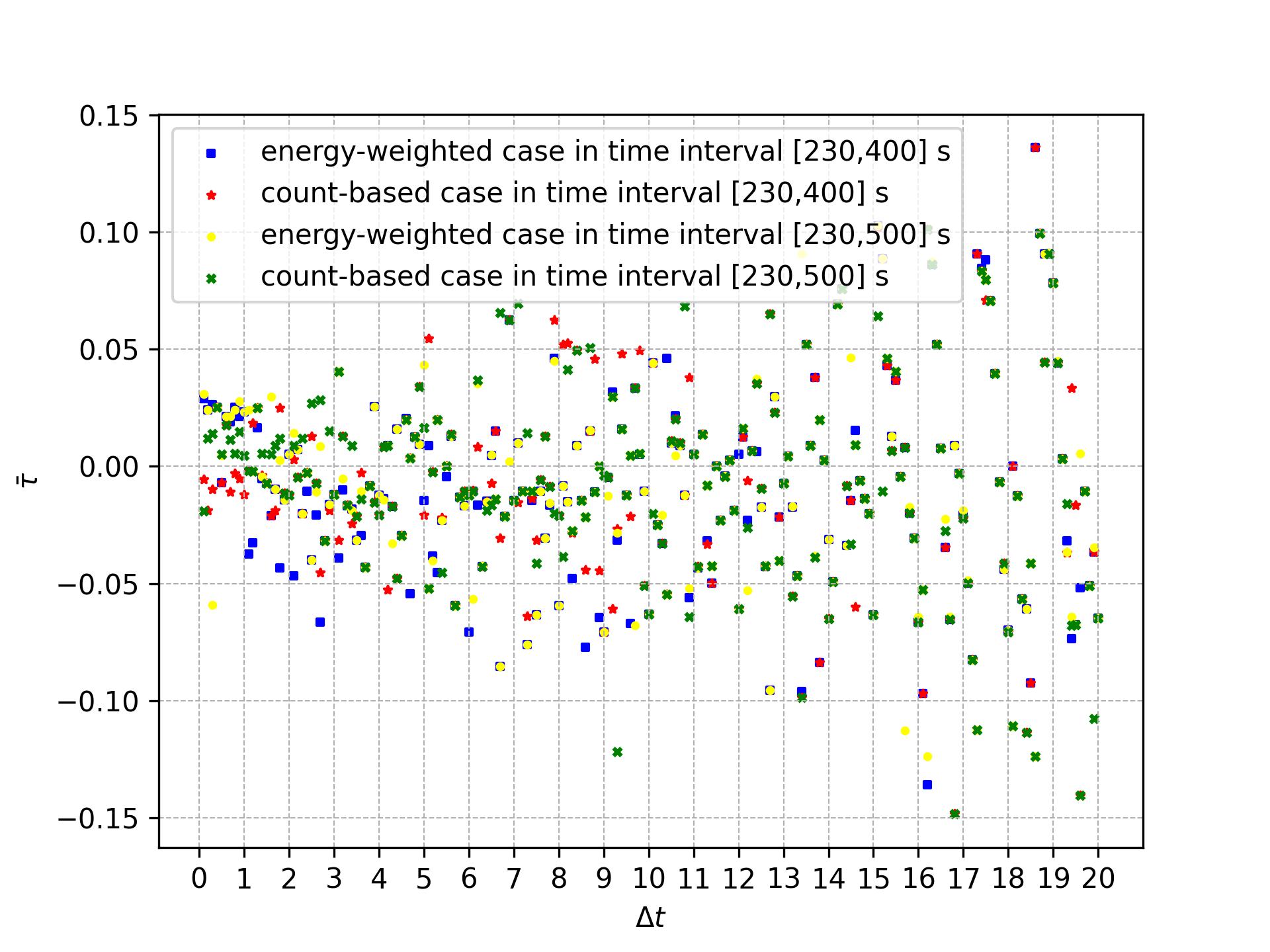} 
        %\caption{}
    \end{minipage}
    \caption{
    The dependence of the optimal parameter $\bar{\tau}$ on $\Delta t$. Left: 1st-order case; Right:2nd-order case.
    }
    \label{4-3}
\end{figure}

We focus on photons detected within the time interval of [230, 400]~s after the trigger time $T_0$, as this temporal window exhibits the highest photon flux and the dominant population of high-energy events.   
 To assess potential selection biases, we also examine an extended time interval of [230,500]~s. 
The variations of Shannon entropy  with the \ac{1st-order} and \ac{2nd-order} LIV parameters are illustrated in Figs.~\ref{4-1} and \ref{4-2}, respectively. 
The red (blue) and green (yellow) curves represent the results derived using the count-based (energy-weighted) profile in the [230,400] s and [230,500] s intervals, respectively. 
Furthermore, the dependence of the optimal parameter $\bar{\tau}$ on $\Delta t$ for both profile types, temporal selections, and LIV scenarios is shown in Fig.~\ref{4-3}. 
Note that our method is different from that used for WCDA photons in \cite{xi2025constraintslorentzinvarianceviolation}. 
They subtract the Shannon entropy of the KM2A and WCDA photons from that of the WCDA photons to obtain the final entropy, while we  use the joint entropy of both the KM2A and WCDA photons. 
Furthermore, we use the fixed time bin width in our analysis, while the Knuth binning method is adopted in \cite{xi2025constraintslorentzinvarianceviolation}. 

Our results show that the energy-weighted profile more effectively characterizes the correlation between Shannon entropy and the LIV parameters compared to the count-based profile, thereby enhancing the precision of parameter estimation. 
We also observe that the different temporal selections for photons have little impact on the results for the 1st-order LIV parameter derived via the energy-weighted profile, as well as for the 2nd-order LIV parameter using both profiles. However, the choice of temporal binning significantly influences the results. 
Excessively narrow time bins (too small $\Delta t$) yield significant oscillations. This artifact arises because photons with almost identical arrival times may be assigned to different time bins under varying parameter configurations, thereby introducing fluctuations in the calculations of Shannon entropy. Conversely, overly wide time bins also degrade accuracy due to information loss. When multiple photons are grouped within a single bin, their individual temporal signatures are effectively averaged, thereby obscuring distinctive features. This effect becomes increasingly severe with wider bins, systematically increasing errors in parameter estimation.

Given the inherent energy resolution limitations of ground-based detector arrays, we incorporate uncertainties in photon energy into our analysis through a Monte Carlo approach. 
We focus on photons in time interval [230,400] s. 
For each photon detected by the \ac{KM2A} detector, we generate randomized energy values following a composite probability distribution comprising two truncated normal distributions. These two distributions are centered on the mean value of the observed energy, with standard deviations corresponding to the reported left and right energy uncertainties \cite{2023_2}, respectively. All generated energies are restricted to exceed 1 TeV. For photons detected by the \ac{WCDA} within the $N_{\text{hit}}$ bin of [30,33], we generate energies using a single normal distribution with a mean value of 0.35 TeV and a standard deviation of 0.35 TeV that reflects the associated uncertainty. The generated energies are restricted to lie between 0.1 TeV and 0.7 TeV. 

\begin{figure}[htbp]
    \centering
    \begin{minipage}{0.48\textwidth}
        \centering
        \includegraphics[width=\linewidth]{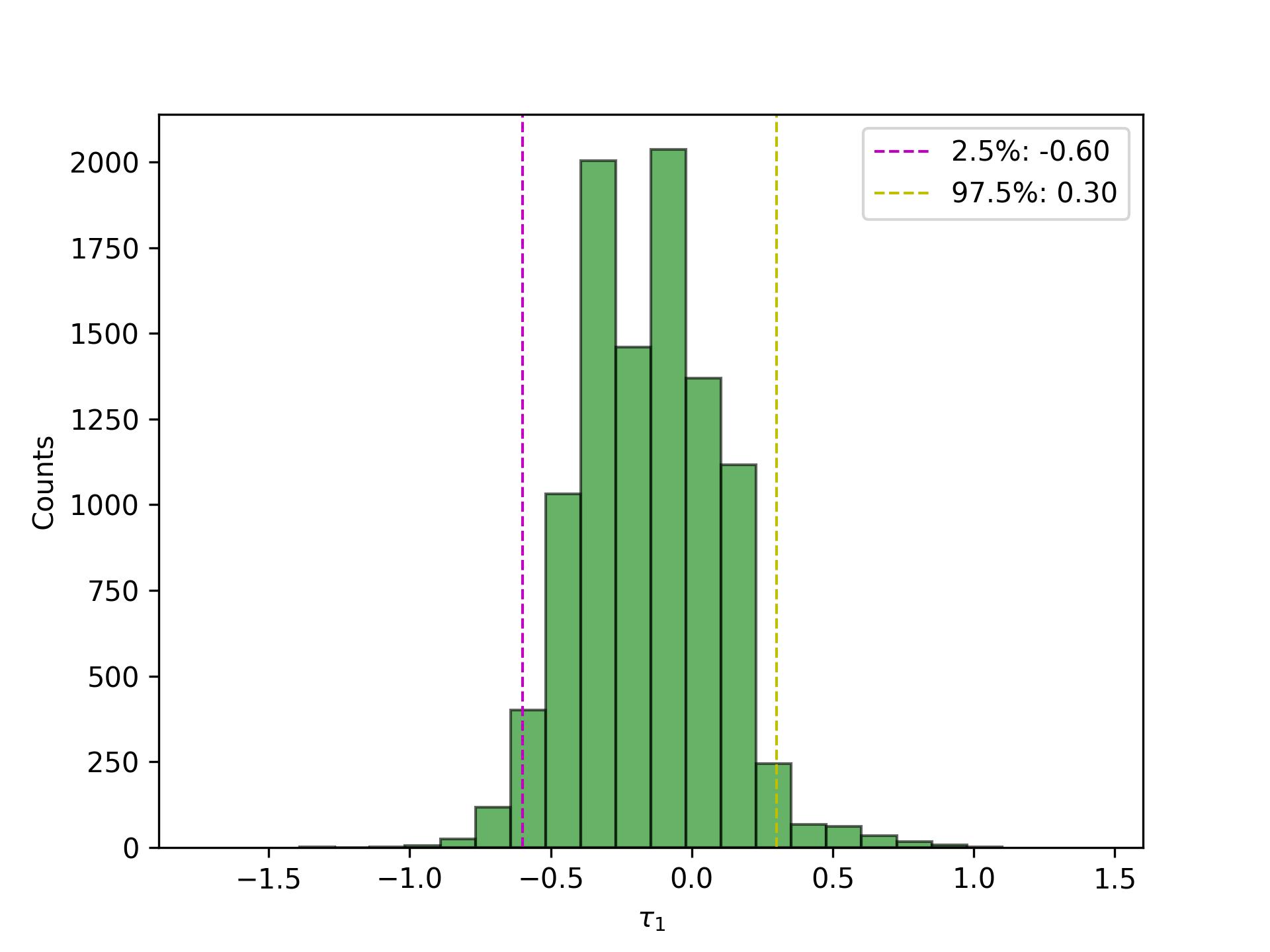}
        %\caption{}
    \end{minipage}
    \hfill
    \begin{minipage}{0.48\textwidth}
        \centering
        \includegraphics[width=\linewidth]{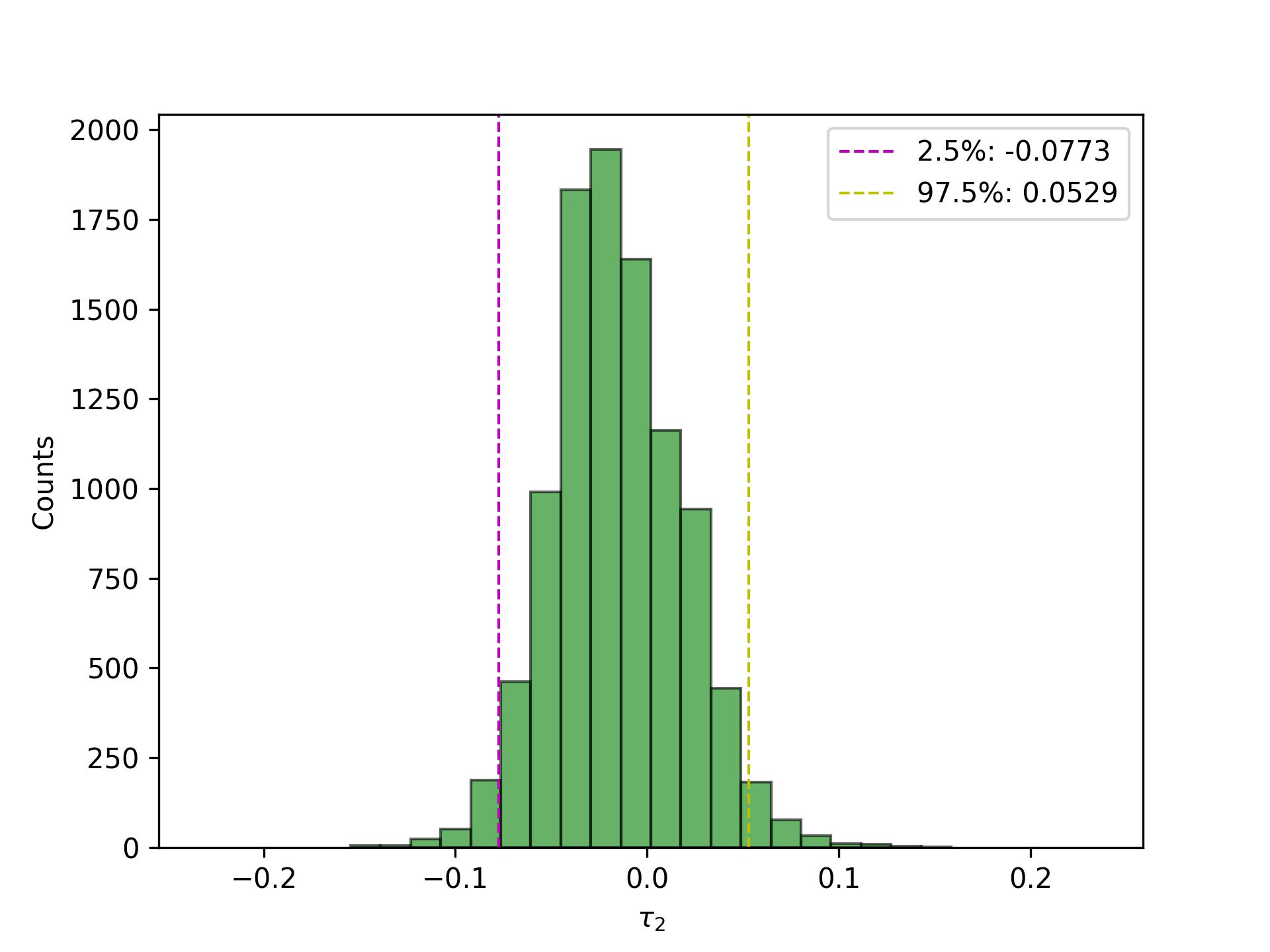}
        %\caption{}
    \end{minipage}
    \caption{
    Distributions of $\bar{\tau}_1$(left) and $\bar{\tau}_2$(right) with the time bin width $\Delta t = 1.6$ s.
    }
    \label{4-?}
\end{figure}

\begin{figure}[htbp]
    \centering
    \begin{minipage}{0.48\textwidth}
        \centering
        \includegraphics[width=\linewidth]{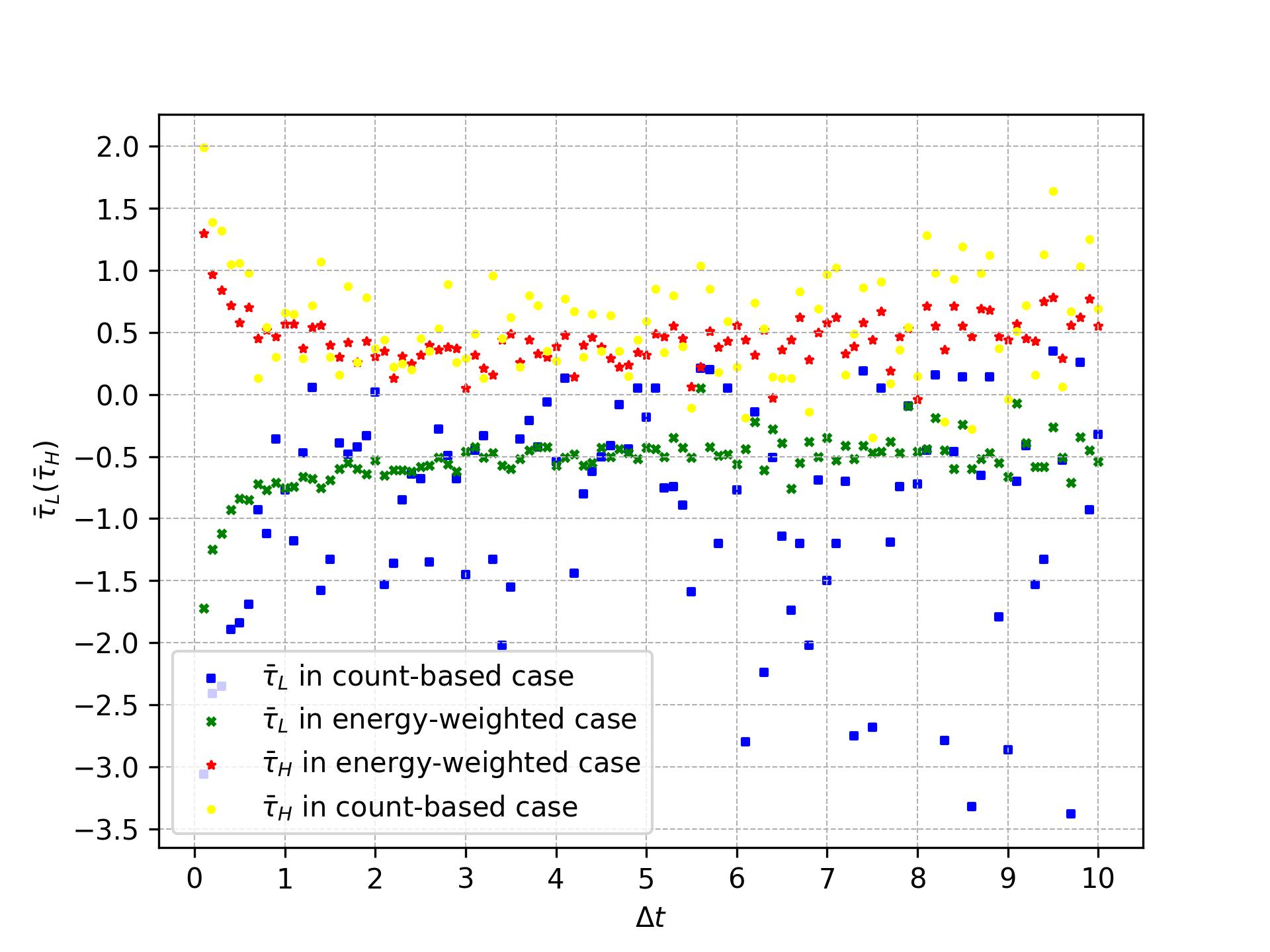}
        %\caption{}
    \end{minipage}
    \hfill
    \begin{minipage}{0.48\textwidth}
        \centering
        \includegraphics[width=\linewidth]{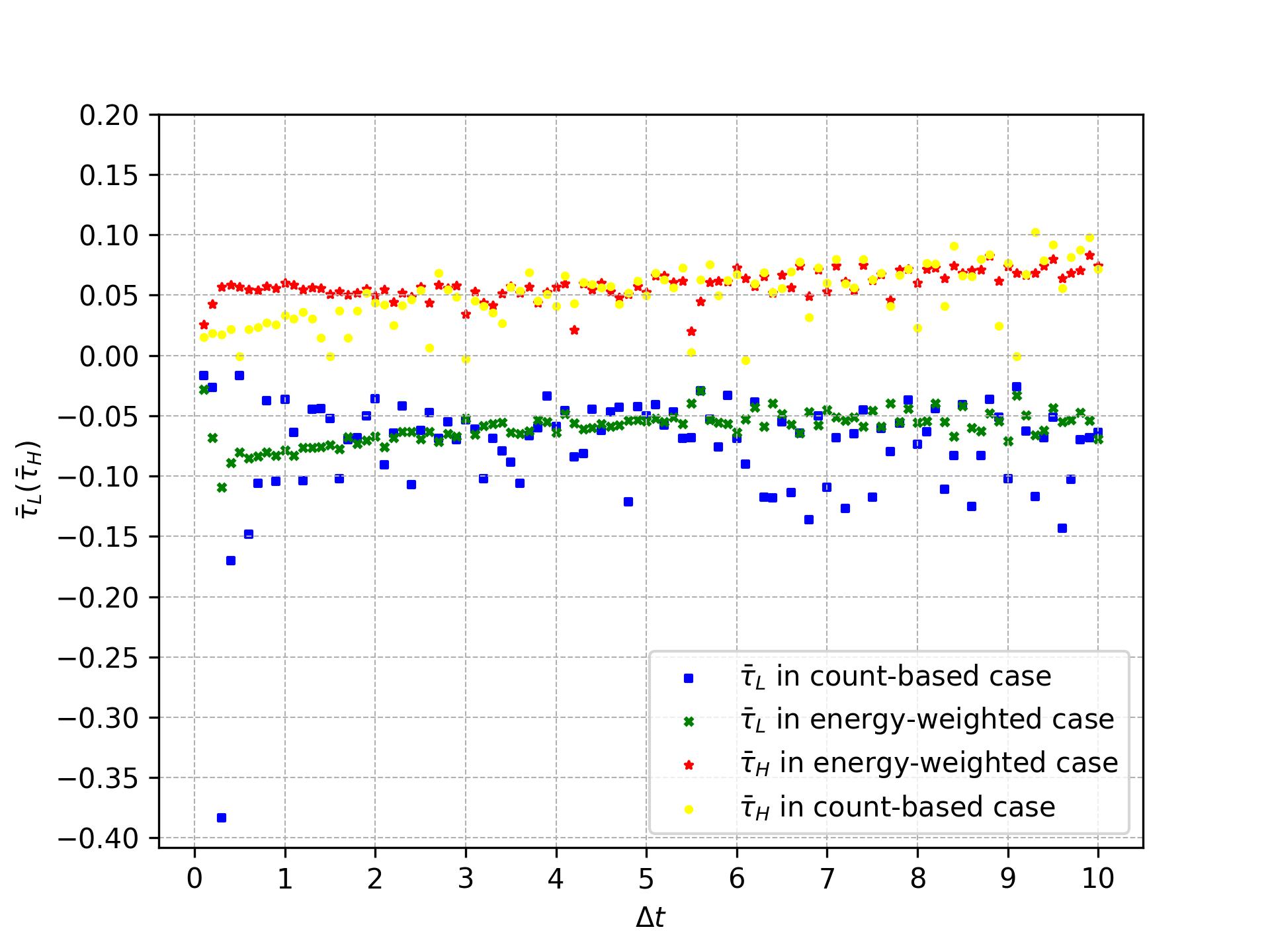}
        %\caption{}
    \end{minipage}
    \caption{ The dependence of the percentile parameters $\bar{\tau}$ on $\Delta t$ for the 1st-order (left) and 2nd-order (right) LIV scenario. The blue (green) and yellow (red) points represent $\bar{\tau}_L$ and $\bar{\tau}_H$ for the count-based (energy-weighted) profile, respectively.
    }
    \label{4-4}
\end{figure}

We generate a set of $10^4$ simulated datasets to statistically assess the impact of energy measurement uncertainties on the determination of the LIV parameter. For each dataset, we identify the optimal LIV parameter $\bar{\tau}$ corresponding to the minimum Shannon entropy. This procedure yields a probability distribution of $\bar{\tau}$ values that accounts for the systematic uncertainty introduced by the energy reconstruction. 
Fig.~\ref{4-?} shows the distributions of $\bar{\tau}_1$ and $\bar{\tau}_2$ with a time bin width $\Delta t = 1.6$ s. 
From these distributions, we extract the 2.5th and 97.5th percentiles of $\bar{\tau}$, denoted as $\bar{\tau}_{L}$ and $\bar{\tau}_{H}$, respectively, which can be converted into constraints on the quantum gravity energy scale $E_{\rm{QG}}$. As noted previously, the determination of $\bar{\tau}$ exhibits a significant dependence on the chosen time bin width $\Delta t$. To systematically evaluate this effect, we perform our analysis across a range of bin widths $\Delta t$ from 0.1 s to 10 s in 0.1 s increments, calculating $\bar{\tau}_{L}$ and $\bar{\tau}_{H}$ for each time bin width.
Fig.~\ref{4-4} illustrates the dependence of these confidence bounds for both the energy-weighted and count-based profiles on $\Delta t$ for the \ac{1st-order} and \ac{2nd-order} LIV effects, respectively. 
Our results show that both $\bar{\tau}_{L}$ and $\bar{\tau}_{H}$ for both the \ac{1st-order} and \ac{2nd-order} LIV scenarios exhibit less dispersion for the energy-weighted profile compared to the count-based profile. 
Furthermore, we observe a long tail at small $\Delta t$ values and large dispersion at large $\Delta t$ values for the energy-weighted profile. These features confirm our earlier discussions regarding the effects of temporal binning. Our analysis suggests an optimal bin width in the range of approximately 1 to 6 seconds, where the parameter estimates exhibit enhanced stability and precision.

\begin{figure}[htbp]
    \centering
    \begin{minipage}{0.48\textwidth}
        \centering
        \includegraphics[width=\linewidth]{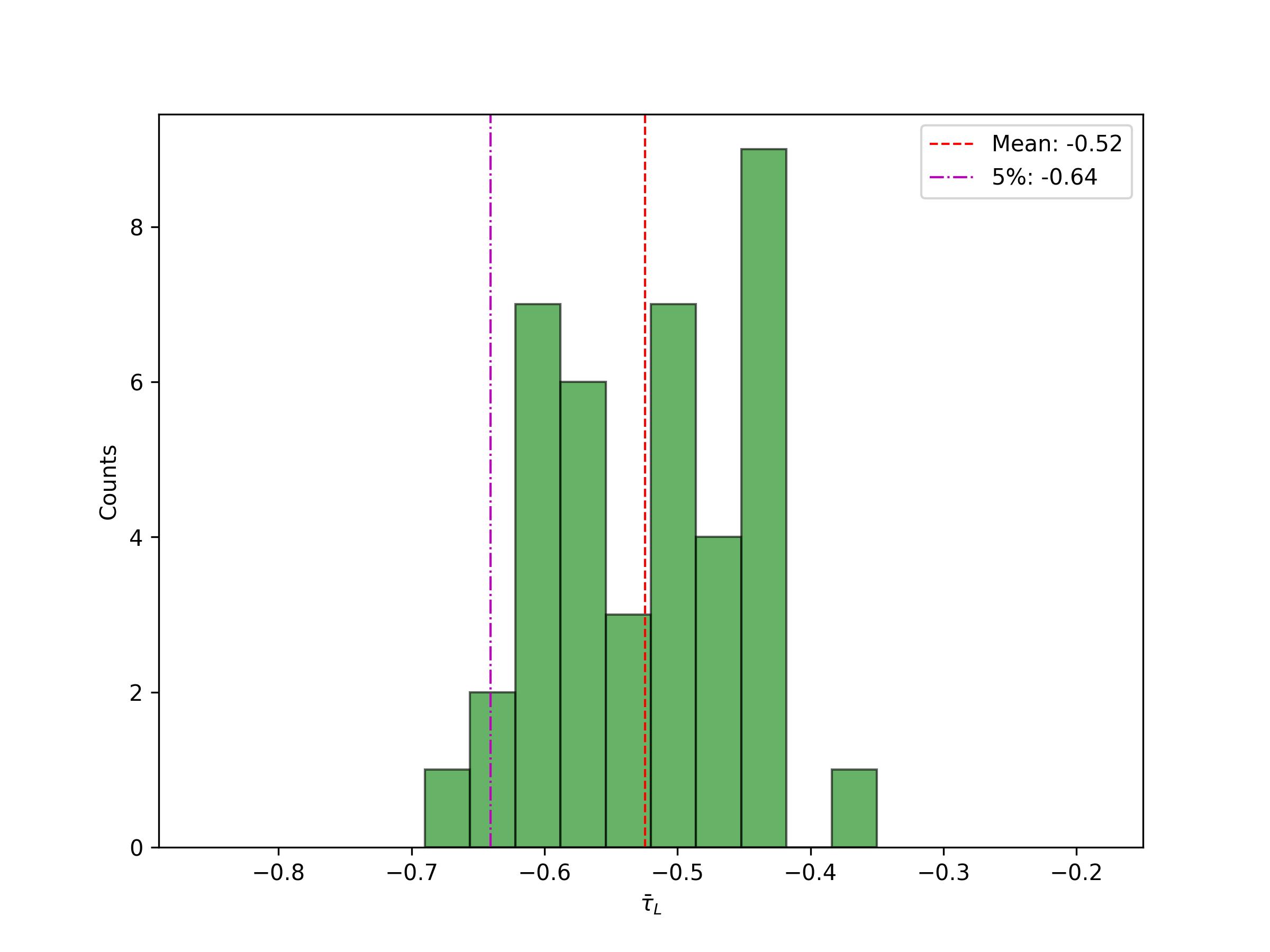}
        %\caption{}
    \end{minipage}
    \hfill
    \begin{minipage}{0.48\textwidth}
        \centering
        \includegraphics[width=\linewidth]{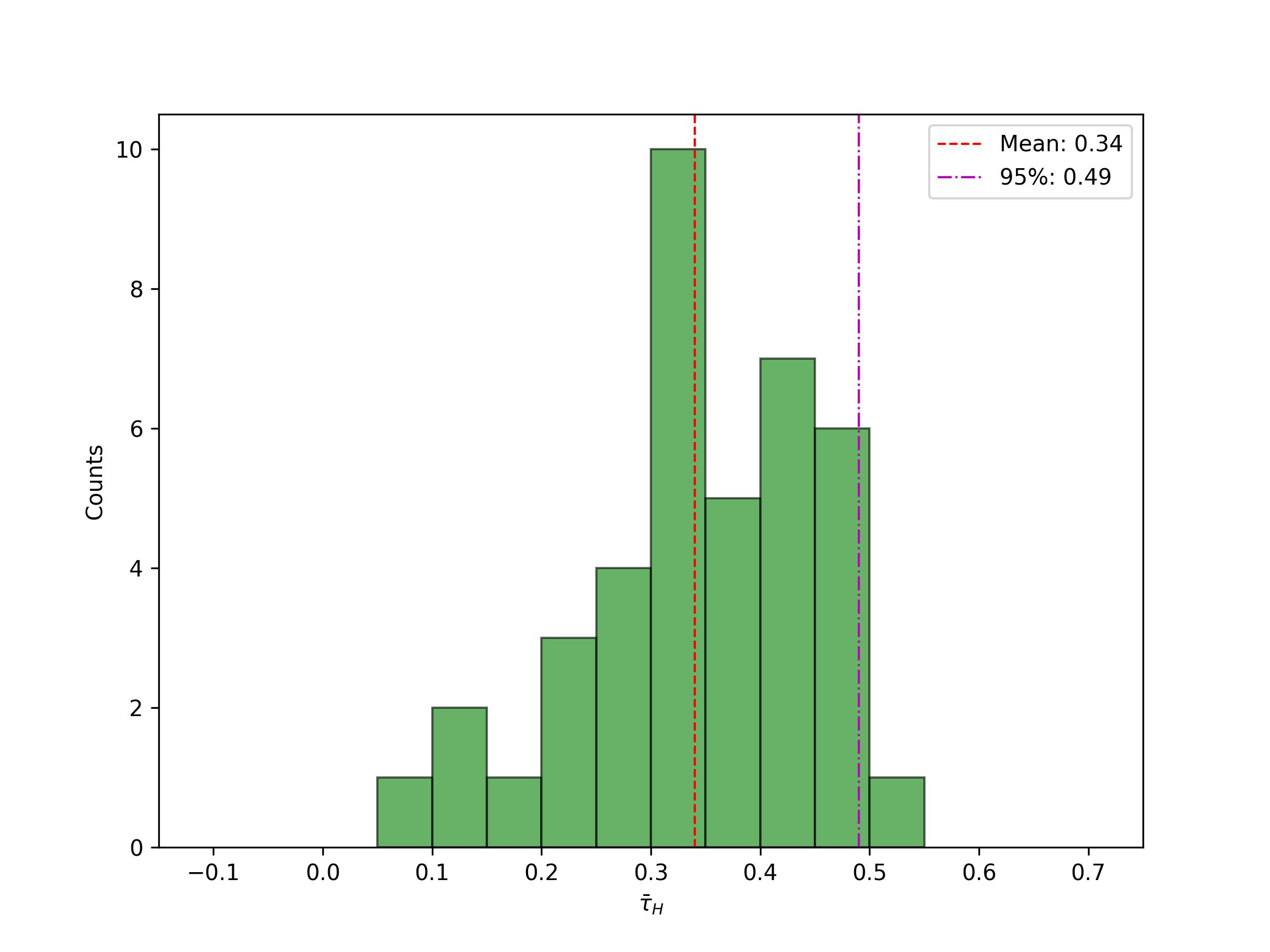}
        %\caption{}
    \end{minipage}
    \caption{
    The distributions of $\bar{\tau}_{L}$ (left) and $\bar{\tau}_{H}$ (right) for the \ac{1st-order} \ac{LIV} effects.
    }
    \label{4-5}
\end{figure}

\begin{figure}[htbp]
    \centering
    \begin{minipage}{0.48\textwidth}
        \centering
        \includegraphics[width=\linewidth]{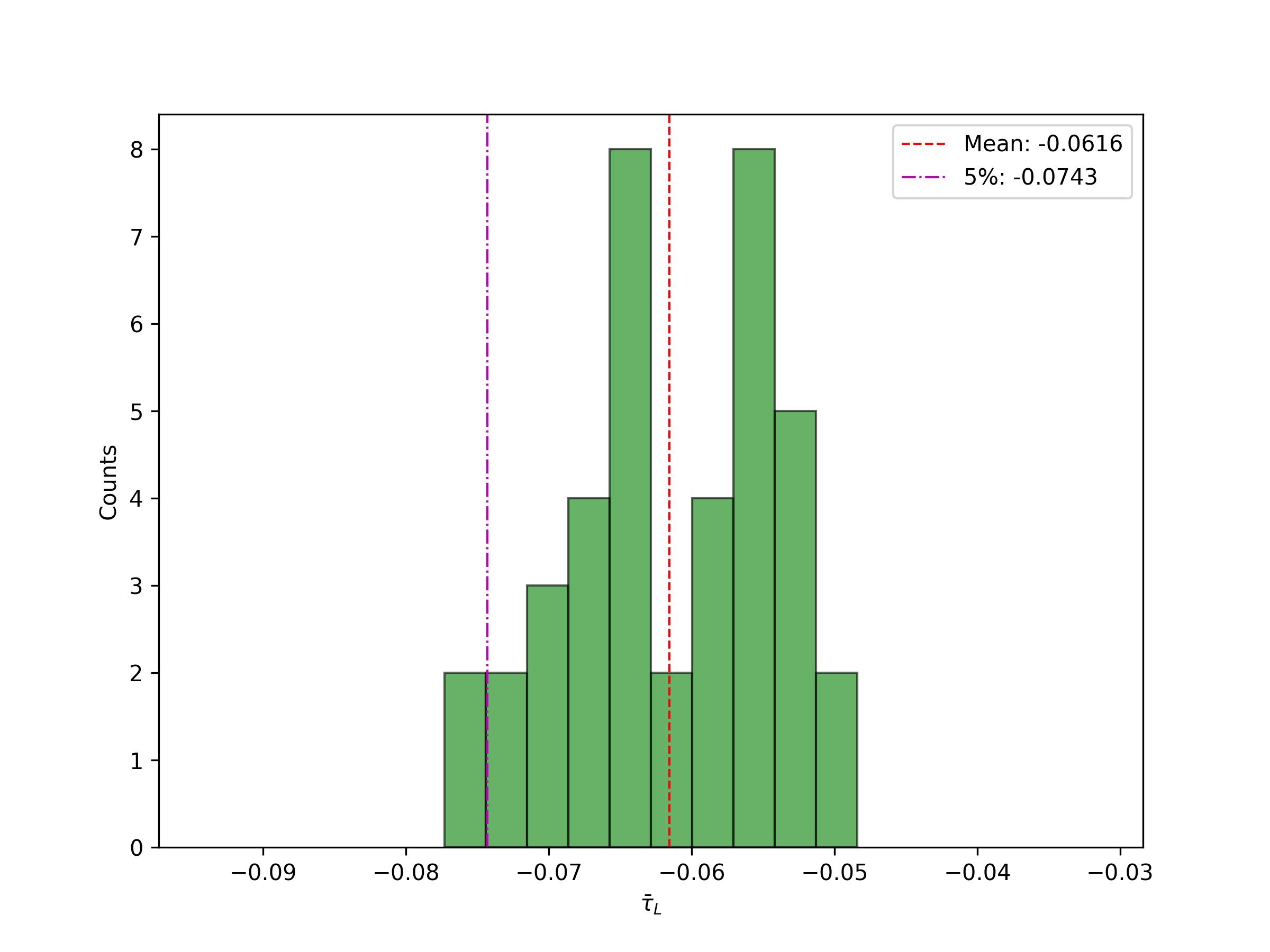}
        %\caption{}
    \end{minipage}
    \hfill
    \begin{minipage}{0.48\textwidth}
        \centering
        \includegraphics[width=\linewidth]{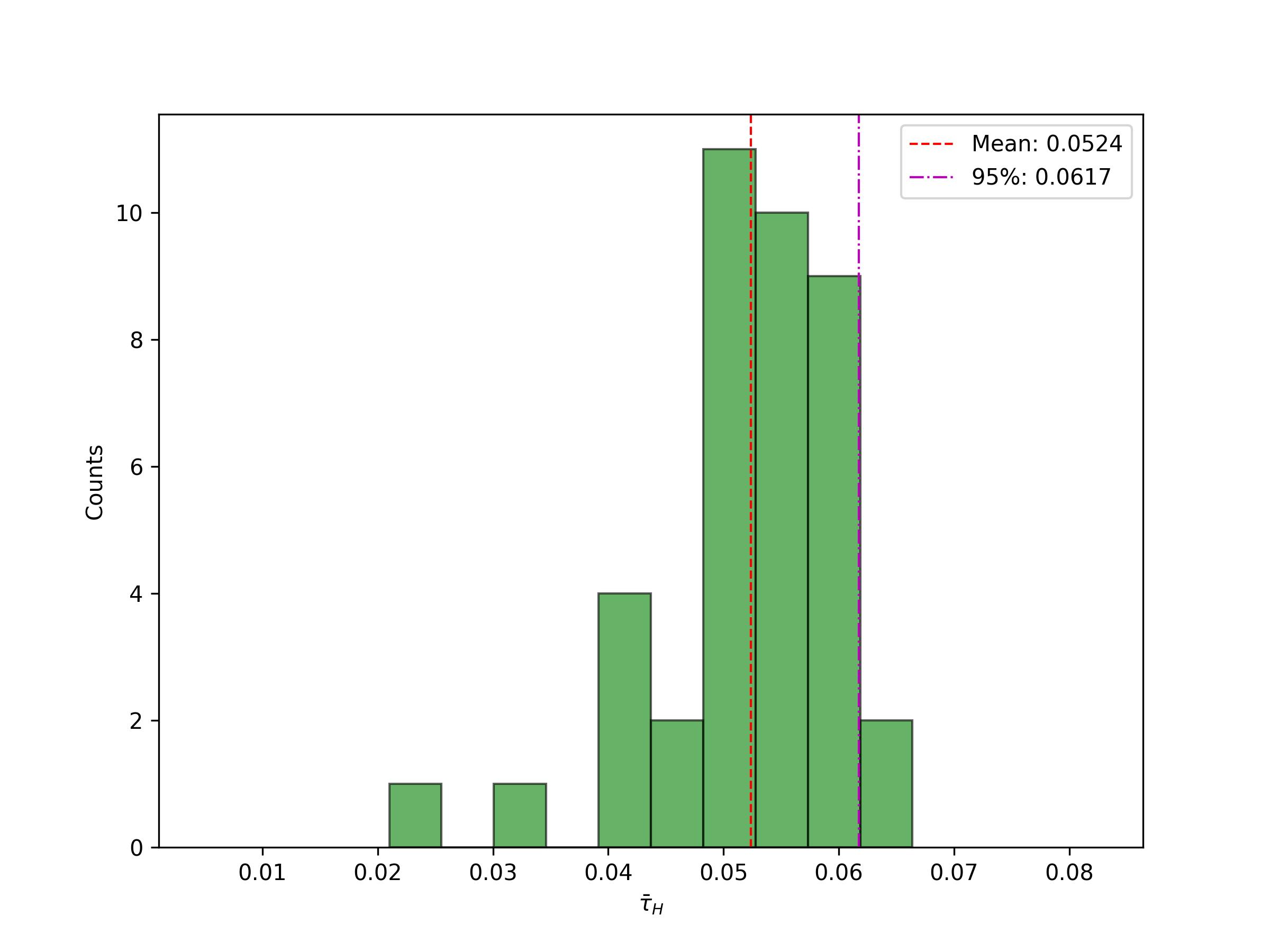}
        %\caption{}
    \end{minipage}
    \caption{
    The same as Fig.~\ref{4-5}, but for the \ac{2nd-order} \ac{LIV} effects. 
    }
    \label{4-6}
\end{figure}

Since the energy-weighted profile shows greater efficiency than the count-based profile, we restrict our subsequent analysis to the energy-weighted profile. 
The final results are shown in Tab.~\ref{result}. 
In our \ac{1st-order} LIV analysis, we examine time bin widths spanning 1.5 to 5.4 seconds in increments of 0.1 seconds. Fig.~\ref{4-5} illustrates the distributions of $\bar{\tau}_{L}$ (left panel) and $\bar{\tau}_{H}$ (right panel). From these distributions, we calculate both the mean values ("stringent" in Tab.~\ref{result}) and $95\%$ confidence intervals ("conservative" in Tab.~\ref{result}) for both parameters. Using the mean values of $\bar{\tau}_{L}$ and $\bar{\tau}_{H}$ as thresholds, we establish the $95\%$ confidence-level constraints on the quantum gravity energy scale $E_{\rm{QG}}$, yielding results of $E_{\rm{QG}}/10^{19}~\text{GeV}> 13.8$ and $E_{\rm{QG}}/10^{19}~\text{GeV}> 21.1$ for the superluminal ($\mathcal{S}=-1$) and subluminal ($\mathcal{S}=+1$) cases, respectively. By employing the boundaries of the $95\%$ confidence interval of the $\bar{\tau}_{L}$ (left panel) and $\bar{\tau}_{H}$ (right panel) distributions as thresholds, we obtain more loose constraints: $E_{\rm{QG}}/10^{19}~\text{GeV}> 11.2$ and $E_{\rm{QG}}/10^{19}~\text{GeV}> 14.6$ the respective cases. The \ac{2nd-order} LIV analysis follows an analogous procedure, examining the same bin widths from 1.5 to 5.4 seconds in increments of 0.1 seconds. The corresponding parameter distributions are shown in Fig.~\ref{4-6}. Based on the mean values ($95\%$ intervals) of the $\bar{\tau}_{L}$ and $\bar{\tau}_{H}$ distributions, we impose the $95\%$ constraints on $E_{\rm{QG}}$ as $E_{\rm{QG}}/10^{11}~\text{GeV}> 13.7$ (12.5) and $E_{\rm{QG}}/10^{11}~\text{GeV}> 14.9$ (13.7) for the superluminal and subluminal cases, respectively. These results show agreement with previous analyses of GRB 221009A reported in the literature \cite{Yang:2023kjq}.

\begin{table}[htbp] 
    \centering 
    \caption{} 
    \label{result} 
\begin{tabular}{|c|c|c|c|c|}
\hline
 & \multicolumn{2}{c|}{$S = -1$} & \multicolumn{2}{c|}{$S = +1$} \\
 \hline
 & stringent & conservative & stringent & conservative \\
\hline
$\tau_1 \cdot \text{ TeV} / \rm{ s}$ & 0.52 & 0.64 & 0.34 & 0.49 \\
\hline
$\tau_2 \cdot \text{ TeV}^2 / \rm{ s}$ & 0.0616 & 0.0743 & 0.0524 & 0.0617 \\
\hline
$E_{\text{QG},1} / 10^{19}\text{ GeV}$ & 13.8 & 11.2 & 21.1 & 14.6 \\
\hline
$E_{\text{QG},2} / 10^{11}\text{ GeV}$ & 13.7 & 12.5 & 14.9 & 13.7 \\
\hline
\end{tabular}
\end{table}

\section{Conclusion}\label{sec:Conclusion}

In this study, we employ the DisCan algorithm to investigate potential \ac{LIV} effects in the \ac{LHAASO} observations of GRB221009A. Utilizing the energies and arrival times of the photons, we calculate the Shannon entropy as a function of the LIV parameter $\tau$ and determine the optimal value $\bar{\tau}$ corresponding to the entropy minimum. To account for uncertainties arising from experimental energy resolution, we generate multiple realizations of photon energies through Monte Carlo sampling, obtaining a probability distribution for $\bar{\tau}$ from which the percentile-based constraints on LIV parameters can be derived. In the analysis, we consider the impact of time binning by examining datasets with varying time bin widths. For the \ac{1st-order} LIV effects, we establish $95\%$ confidence-level constraints on the quantum gravity energy scale $E_{\rm{QG}}$ of $E_{\rm{QG}}/10^{19}~\text{GeV}> 13.8$ and $E_{\rm{QG}}/10^{19}~\text{GeV}> 21.1$ for the superluminal and subluminal cases, respectively. The constraints for the \ac{2nd-order} LIV effects are $E_{\rm{QG}}/10^{11}~\text{GeV}> 13.7$ and $E_{\rm{QG}}/10^{11}~\text{GeV}> 14.9$ for the respective cases.

\acknowledgments

This work is supported by the National Natural Science Foundation of China under grant No. 12175248.

% Bibliography

%% [A] Recommended: using JHEP.bst file
%% \bibliographystyle{JHEP}
%% \bibliography{biblio.bib}

%% or
%% [B] Manual formatting (see below)
%% (i) We suggest to always provide author, title and journal data or doi:
%% in short all the informations that clearly identify a document.
%% (ii) please avoid comments such as ``For a review'', ``For some examples",
%% ``and references therein'' or move them in the text. In general, please leave only references in the bibliography and move all
%% accessory text in footnotes.
%% (iii) Also, please have only one work for each \bibitem.

% \begin{thebibliography}{99}

% \bibitem{a}
% Author,
% \emph{Title},
% \emph{J. Abbrev.} {\bf vol} (year) pg.

% \bibitem{b}
% Author,
% \emph{Title},
% arxiv:1234.5678.

% \bibitem{c}
% Author,
% \emph{Title},
% Publisher (year).

% \end{thebibliography}
\bibliography{biblio}
\bibliographystyle{JHEP}
\end{document}